\documentclass[12pt]{article}

\usepackage{geometry}
\usepackage[utf8]{inputenc}
\usepackage[T1]{fontenc}
\usepackage{microtype}
\usepackage{graphicx}
\usepackage{authblk}
\usepackage[hyphens]{url}
\usepackage{hyperref}
\hypersetup{colorlinks=true, linkcolor=blue, citecolor=blue, urlcolor=blue}
\usepackage{caption}
\usepackage{subcaption}
\usepackage{float}
\usepackage{amsmath}
\usepackage{amssymb}
\usepackage{booktabs}
\usepackage{multirow}
\usepackage{array}
\usepackage{tabularx}
\usepackage{longtable}
\usepackage{enumitem}
\usepackage{xcolor}
\usepackage{tcolorbox}
\tcbuselibrary{skins,breakable}
\usepackage{listings}
\usepackage{soul}
\usepackage{fancyvrb}
\usepackage{algorithm}
\usepackage{algpseudocode}
\usepackage{pdflscape}
\usepackage{rotating}
\usepackage{newfloat}
\usepackage{pgfplots}
\pgfplotsset{compat=1.18}
\usepackage{cite}
\newcommand{\OurDatabase}{ScaleMOF }

\definecolor{keywordcolor}{RGB}{0,0,180}
\definecolor{commentcolor}{RGB}{0,128,0}
\title{\textbf{Predicting Scale-Up Plausibility of Metal-Organic Framework
Syntheses}}

\author[1,$\dagger$]{Peter Walther}
\author[1,3,$\dagger$]{Hongrui Sheng}
\author[1,4,$\dagger$]{Xinxin Liu}
\author[1,2]{Bin Feng}
\author[1]{Reid Coyle}
\author[1]{Xinhua Yan}
\author[1]{Kyle Smith}
\author[1]{Harrison Kayal}
\author[1,5]{Shyam Chand Pal}
\author[1,2,*]{Zhiling Zheng}

\affil[1]{Department of Chemistry, Washington University, St.\ Louis, MO 63130, USA.}
\affil[2]{Institute of Materials Science \& Engineering, Washington University,
St.\ Louis, MO 63130, USA.}
\affil[3]{Department of Chemistry, Fudan University, Shanghai 200438, China.}
\affil[4]{Department of Computer and Information Science, University of
Pennsylvania, PA 19104, USA.}
\affil[5]{College of Chemistry and Materials Science, Fujian Normal University,
Fuzhou 350117, China.}

\affil[$\dagger$]{These authors contributed equally.}
\affil[*]{Corresponding author. Email: z.z@wustl.edu}

\date{}

\begin{document}
\maketitle

\begin{abstract}
Scalable synthesis remains the gate between MOF discovery and industrial
deployment, as scale-up know-how is fragmented across disparate reports. We
introduce ScaleMOF, a literature-mined dataset and a
positive-unlabeled learning strategy that fine-tunes large language models.
Achieving 93.5\% accuracy, this proof-of-concept serves as a
literature-grounded ranking tool prioritizing plausible scale-up candidates.
\end{abstract}

Metal-organic frameworks (MOF) have, hitherto, benefitted from the expansive
design space opened by reticular chemistry,\cite{ref01,ref02,ref03,ref04} and
the literature now measures progress by the tens of thousands of
structures,\cite{ref02,ref05,ref06,ref07,ref08} yet the path from crystalline
discovery to industrial material remains slow and not easily transferable across
scales.\cite{ref09,ref10,ref11,ref12} A skilled chemist may progress from
single-crystal synthesis to milligram production, then to sub-gram, gram, and
eventually kilogram or larger batches;\cite{ref13,ref14,ref15,ref16,ref17,ref18}
however, this intuition is hard to be generalized as it is distributed across
disparate synthesis records. Considering the rising industrial interest in MOFs
and the growing proximity between discovery and
application,\cite{ref10,ref19,ref20,ref21,ref22} it is increasingly important
to ask whether scale-up plausibility itself can be predicted from the early
synthesis record.

Herein, we report a data-centric workflow that connects the latent synthesis
knowledge embedded in the literature to the prediction of scale-up plausibility
for MOFs. Our workflow contains a large language model (LLM) agent for data
mining, and the resulting experimental scale-up MOF dataset, which we termed
\OurDatabase dataset, was used to fine-tune another LLM under
positive-unlabeled (PU) learning\cite{ref23,ref24} to give binary
classification for reaction procedures. Our central goal is that, given a newly
reported MOF with a small-scale synthesis description, can we predict whether
the protocol exhibits credible potential for gram-scale or larger production,
and can we do so early enough to guide experimental prioritization and industrial
engagement. In our workflow, we combined LLMs with PU learning to construct a
literature-scale workflow for scalable synthesis prediction, ultimately
facilitating the transition from laboratory discovery to scale-up evaluation.

Additionally, we claim that scale-up prediction is fundamentally distinct from
synthesizability prediction and synthesis-condition recommendation. While
synthesizability models concern whether a MOF can be made, scale-up prediction
focuses on if a known synthesis can be reproduced at $\geq 1$\,g batch size.
Such question is irrelevant for mg-scale curiosity-driven syntheses and has no
established answer in prior literature. Unlike synthesis-condition models that
optimize for crystal quality or yield, scale-up plausibility depends on vessel
type, stirring, solvent system simplicity, and modulator use that are largely
invisible to structure-based approaches. Our dataset is the first to explicitly
label scale-up evidence at the protocol level, and our benchmark design (gold:
$P_\mathrm{s}$ vs.\ $N$; deployment: $P_\mathrm{a}$ vs.\ $U$) mirrors the
practical screening workflow of a process chemist rather than a computational
crystal-structure predictor.

At the onset of this study, we reasoned that the literature already contains
the ingredients for such a model, although they are scattered in forms that are
difficult to unify. Many MOFs have been reported both as initial small-scale
solvothermal crystallizations and, later, as gram-scale or kilogram-scale
preparations under related or modified conditions. Conceptually, these linked
reports constitute an apprenticeship archive of synthesis scale-up. To this
regard, we assembled two literature pools from Web of Science to reflect the
asymmetric nature of scale-up evidence.

The scale-up pool, yielding our possible-positive set (denoted $P$), was
collected using the search keywords ``scale-up'', ``gram-scale'', ``kilogram'',
and ``pilot'' in title or abstracts, under ``metal-organic framework''
peer-reviewed literatures between 1995 and 2026, intentionally with broad scope
so that relevant protocols would not be lost at the search stage; this search
yielded 117 candidate paper groups. Meanwhile, the general synthesis pool,
yielding our unlabeled set (denoted $U$), was collected using
``metal-organic framework'' and ``synthesis'' with the same year range. It was
then restricted to solvothermal syntheses containing one principal metal source
and one principal linker, following the method established in prior MOF data
mining study,\cite{ref25,ref26,ref27,ref28,ref29,ref30,ref31,ref32} and this
selection criterion yielded 946 candidate paper groups.

\begin{figure}[htbp]
  \centering
  \includegraphics[width=\textwidth, height=0.85\textheight, keepaspectratio]{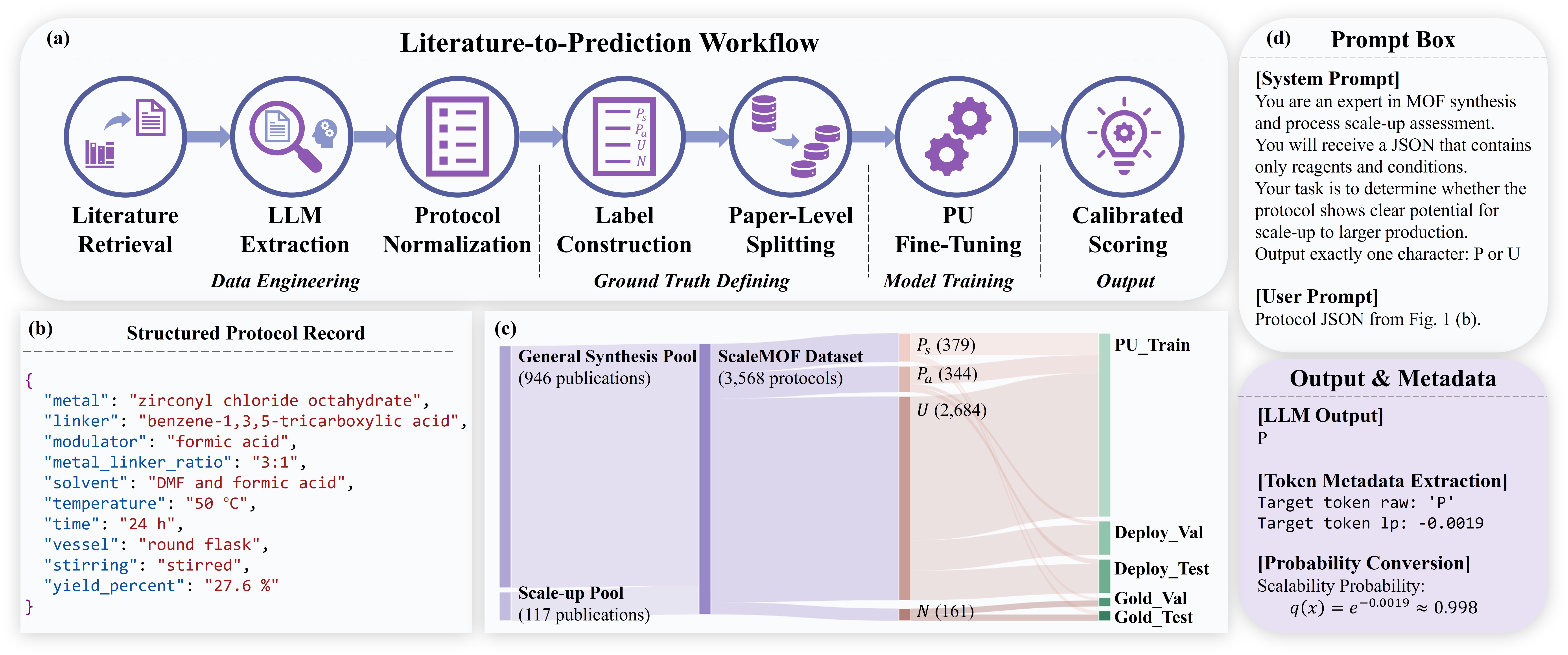}
  \caption{Overview of the \OurDatabase workflow.
    \textbf{(a)} End-to-end pipeline from literature retrieval to scale-up
    prediction.
    \textbf{(b)} Representative JSON reaction card generated by GPT-5.4 from a
    raw synthesis paragraph.
    \textbf{(c)} Data-flow Sankey diagram showing label distribution and split
    allocation.
    \textbf{(d)} Binary classification output of the fine-tuned GPT-4.1
    (\OurDatabase) with representative ``P'' and ``U'' examples.}
  \label{fig:workflow}
\end{figure}

To bridge the gap between these unstructured publications and a machine-readable
dataset (see Supporting Information), we text-mined reaction-level synthesis
records\cite{ref25,ref28} from both the main manuscripts and associated
supporting information files (if presented). The overall workflow is shown in
Fig.~\ref{fig:workflow}. In particular, GPT-5.4 was first deployed as a
data-mining agent; analyzing the full text of a paper, it extracted essential
fields—including the primary metal precursor, organic linker, modulators, solvent
systems, and reaction conditions—yielding the structured protocol records
exemplified in Fig.~\ref{fig:workflow}(b). The subsequent flow of these curated
protocols, label constructions, and data splits is summarized in
Fig.~\ref{fig:workflow}(c) and Supporting Information Section~S1. In addition,
the representative LLM prompts and the resulting binary predictions were
demonstrated in Fig.~\ref{fig:workflow}(d) and Supporting Information,
Section~S3.

While structure identifier (e.g., MOF names and CCDC
number\cite{ref33,ref34}), reagent mass and volume, post-processing conditions
(e.g., solvent washing time and activation temperature), and reaction outcomes
(e.g., isolated product mass and yield), were also considered and extracted
under the same schema, they were optional items and kept blank if the authors
of the selected papers did not explicitly report these values (Supporting
Information, Sections S2.3 \& S2.4). A random 10\% of selected papers from
both corpora were manually verified by comparing with ground truth annotated by
the human expert, giving an extraction accuracy of 97.6\% for key reaction
parameters (precursor, solvent, time, temperature), which is consistent with
the high accuracy of chemical data extraction in LLM-driven data
mining.\cite{ref25,ref27,ref28,ref35,ref36}

The resulting dataset, termed the \OurDatabase dataset, contains 3,568 synthesis
protocols, with labels defined as follows: strong positives
($P_\mathrm{s}$, 379 protocols with explicit scale-up evidence), auxiliary
positives ($P_\mathrm{a}$, 344 protocols in the unlabeled corpus that match
$P_\mathrm{s}$ by MOF name and metal/linker identity, representing small-scale
preparations of a MOF later shown to be scalable), unlabeled ($U$, 2,684
entries containing unknown positives and true negatives), and a small negative
set ($N$, 161 entries) curated by human experts. These negative examples
represent protocols deemed fundamentally unscalable (e.g., restricted strictly
to mg-scale yields or employing extreme, impractical conditions). After
deduplication, the \OurDatabase dataset contains 2,684 unlabeled protocols
(75.2\%), while $P_\mathrm{s}$ and $P_\mathrm{a}$ together contribute 723
positives (20.3\%), and $N$ contributes 161 (4.5\%). A comprehensive chemical
diversity analysis of this curated dataset, including UMAP and t-SNE projections
of the protocol embeddings, product-mass distributions, and synthesis condition
statistics, is provided in the Supporting Information (Figures~S1 and S2).

To prevent data leakage, protocols were split by paper-level group identifiers
in an approximate 70:15:15 ratio so that no single publication contributes to
more than one of the train, validation, or test sets. Crucially, according to
the PU learning strategy, the entire negative set was explicitly excluded from
the paper-level training split and instead allocated exclusively and nearly
equally to the gold-standard validation and test sets.

Consequently, for model training, the PU learning subset contained 2,420
protocols (290 $P_\mathrm{s}$, 234 $P_\mathrm{a}$, and 1,896 $U$) in the
training split. For evaluation, the $N$ examples were used exclusively in the
gold-standard evaluation benchmarks: the validation gold set contains 118
protocols (38 $P_\mathrm{s}$ and 80 $N$), and the test gold set contains 132
protocols (51 $P_\mathrm{s}$ and 81 $N$). Furthermore, the deployment
benchmarks, which simulate realistic literature screening, comprise 448
protocols in the validation deployment set (47 $P_\mathrm{a}$ and 401 $U$) and
450 protocols in the test deployment set (63 $P_\mathrm{a}$ and 387 $U$).

The predictive model was then structured by fine-tuning a base LLM on JSON
representations of these synthesis protocols.\cite{ref27,ref32,ref37,ref38,ref39}
In training, both $P_\mathrm{s}$ and $P_\mathrm{a}$ were mapped to the positive
label P, whereas $U$ remained unlabeled and $N$ was excluded from training.
This construction is based on the assumption that absence of scale-up evidence
in literature does not imply failure, but only incomplete knowledge.
Positive-unlabeled learning is therefore well suited for this classification
task,\cite{ref24} where input is a ``reaction card'' and output is a
single-token ``P'' or ``U'' label, and the complex nature of reagent names and
reaction apparatus make LLMs a good fit for the model shown in
Fig.~\ref{fig:workflow}(d). For the PU correction, we adopted the classical
perspective that the observed positive label in the literature is a biased
sampling of true positives. Consequently, the raw probabilities output by the
model may inherently underestimate a protocol's true scalability, necessitating
a mathematical calibration.

Operationally, we obtained the model's raw score as the token-level probability
of emitting label P, denoted $q(x)$, derived from log probabilities of the
actual output token for an input protocol $x$. In practice, we then estimated
$\hat{c}$, the mean of $q(x)$ over validation positives $\mathcal{V}_{P_s}$ in
the validation gold set, and this factor was computed to be 0.8410. This constant
implies that while the labelled literature captures the majority of scalable
protocols, approximately 16\% of genuinely scalable MOFs remain undocumented
and hidden within the unlabeled corpus. Because industrial screening is
fundamentally a ranking problem, we also applied a calibration layer via Platt
scaling\cite{ref40} on the gold validation set to produce a final score whose
threshold may be selected for a desired trade-off between missed opportunities
and false leads, and the resulting validation threshold was determined to be 0.11.

Yet, we acknowledge that our composite positive set $P$ does not necessarily
satisfy the strict ``Selected Completely At Random'' (SCAR) assumption, which
is a standard PU learning premise indicating that labeled examples must be a
completely unbiased, random subsample of all true positives. Since $P_\mathrm{a}$
is constructed by targeted MOF-identity matching rather than independent random
labelling, it inherently reflects the publication bias of chemical literature.
We therefore treat $\hat{c}$ as a pragmatic calibration factor rather than a
theoretically exact PU correction. Using only $P_\mathrm{s}$ for $\hat{c}$
estimation is deliberately conservative: $P_\mathrm{s}$ records have the most
reliable scale-up evidence, making the average $q(x)$ on this set a stable
lower bound on the true labelling frequency. The consistently strong predictive
performance on both the gold holdout and the deployment benchmark validates this
pragmatic approach empirically, even in the absence of strict theoretical
guarantees.


\begin{figure}[htbp]
  \centering
  \includegraphics[width=0.9\textwidth, height=0.5\textheight, keepaspectratio]%
    {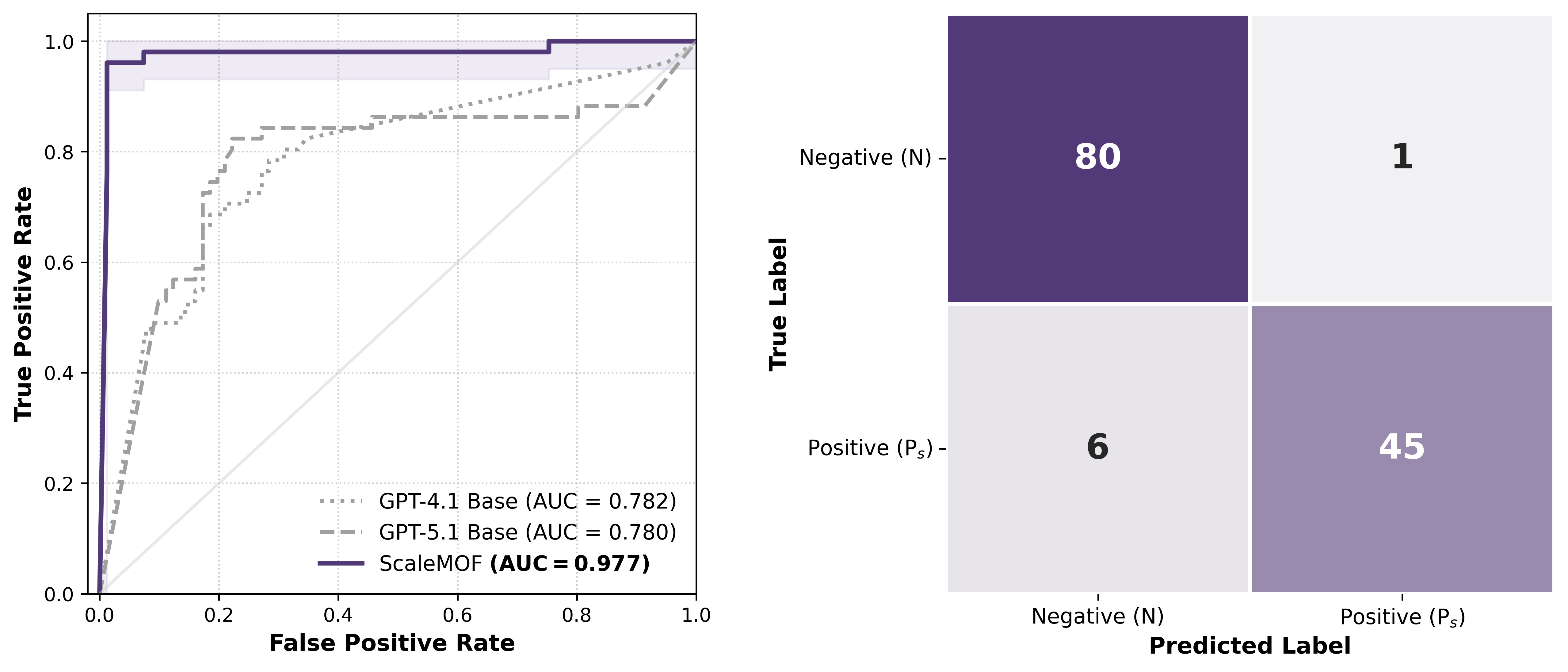}
  \caption{Predictive performance of the fine-tuned LLM on the \OurDatabase
    dataset.
    \textbf{(a)} Receiver operating characteristic (ROC) curves comparing
    \OurDatabase with zero-shot base models. \OurDatabase achieves the highest
    area under the curve (AUC\,$= 0.977$); the shaded region represents the
    bootstrap confidence band.
    \textbf{(b)} Confusion matrix of \OurDatabase at decision threshold
    $\tau = 0.11$ on the gold test set ($n = 132$):
    TP\,$= 45$, TN\,$= 80$, FP\,$= 1$, FN\,$= 6$
    (balanced accuracy 93.5\%, F1 92.8\%).}
  \label{fig:roc_cm}
\end{figure}

\begin{figure}[htbp]
  \centering
  \includegraphics[width=\textwidth, height=0.45\textheight, keepaspectratio]%
    {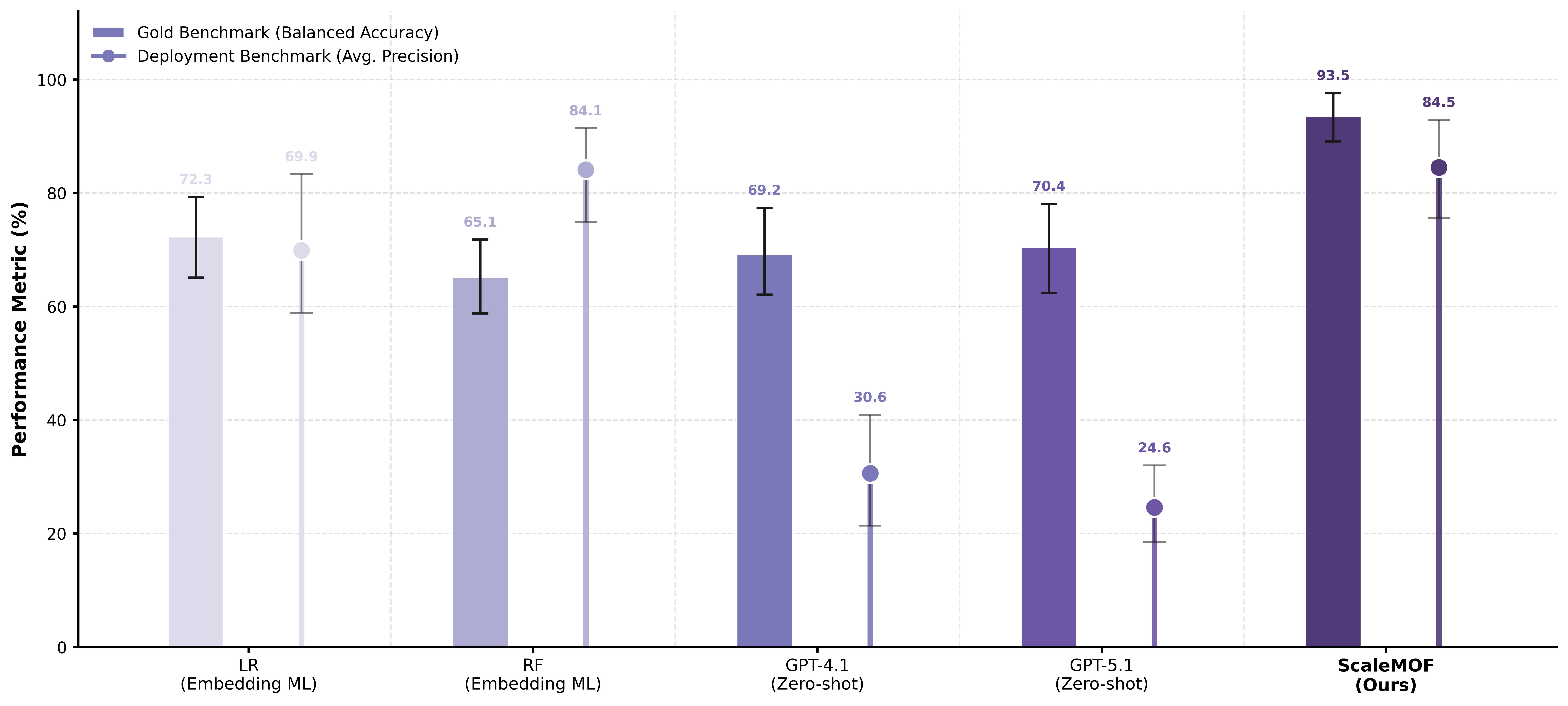}
  \caption{Performance comparison of \OurDatabase with baseline models.
    Left bars: Gold Benchmark balanced accuracy ($P_\mathrm{s}$ vs.\ $N$,
    $n = 132$) with 95\% bootstrap CI.
    Right lollipops: Deployment Benchmark average precision ($P_\mathrm{a}$
    vs.\ $U$, $n = 450$) with 95\% bootstrap CI. Embedding ML baselines
    (LR and RF on 1,536-dim text embeddings) are included to address
    information-matching concerns.
    The fine-tuned \OurDatabase consistently and substantially outperforms
    all baselines across both metrics.}
  \label{fig:comparison}
\end{figure}


Upon the model fine-tuning, the first evaluation was designed to test whether
the model can distinguish explicit scale-up protocols from unlikely ones under
controlled conditions. On the gold benchmark, which compares $P_\mathrm{s}$
against $N$, the validation set contained 118 protocols and the test set 132
protocols. Using the raw observed-positive score alone, the balanced accuracy
was 89.6\%. After PU correction and calibration, balanced accuracy increased to
93.5\%, with a F1 score of 92.8\% and Matthews correlation
coefficient\cite{ref41,ref42} of 88.9\%, while ROC-AUC reached 97.7\%
(Fig.~\ref{fig:roc_cm}(a); Supporting Information, Section~S4.1), which shed
light on a key point that the remaining errors are limited and balanced rather
than concentrated in one failure mode. Bootstrap analysis further supported the
robustness of this performance, with the 95\% confidence interval for balanced
accuracy spanning 89.1 to 97.6\% and that for F1 spanning 87.4 to 97.4\%.

A more demanding and, in our view, more useful test asks whether the model can
surface promising small-scale protocols from the large body of ordinary
literature where scalability is not explicitly declared. For this reason, we
evaluated deployment performance on $P_\mathrm{a}$ versus $U$. The validation
deployment set contained 448 protocols, whereas the held-out test deployment set
contained 450 protocols composed of 63 ancillary positives and 387 unlabeled
candidates. On this test, the final calibrated model achieved an ROC-AUC of
96.9\%. Additionally, precision at the top of the ranking was strong, reaching
70.0\% at 10. Because a researcher typically screens a family of related
syntheses within a single publication, we also evaluated the model's localized
ranking power. This paper-level evaluation is restricted to test papers that
contain both ancillary positives ($P_\mathrm{a}$) and unlabeled protocols ($U$)
within the test partition, the minimum condition for a meaningful within-paper
ranking. On this granular test, the model achieved a top-3 hit rate of 100.0\%
across 5 held-out test papers. It is envisioned that such a ranking mode is
especially valuable for industrial triage, where one rarely needs a model to
certify every protocol; rather, one needs it to move the most plausible
candidates to the front of the queue.

We next asked whether fine-tuning truly adds chemical discrimination beyond
that available in a general-purpose model (Fig.~\ref{fig:comparison}). On the
gold test benchmark, the balanced accuracy of the fine-tuned model (93.5\%) is
significantly higher than those of the base models at 78.8\% (GPT-5.1) and
68.7\% (GPT-4.1), respectively. On the more practically relevant deployment
test, fine-tuning widened the margin further as average precision rose to 84.5\%,
whereas the base model only reached 24.6\% (GPT-5.1) to 30.8\% (GPT-4.1).
Even at the paper level, where untuned models retained some within-document
ranking ability, the fine-tuned model still led with a mean average precision
of 87.4\% and a top-1 hit rate of 80.0\%, illustrating high utility for
down-selection in an industrial discovery pipeline.

Notably, this predictive power also substantially eclipses traditional machine
learning baselines (Supporting Information, Section~S4.1), which struggle to
capture the nuanced, contextual heuristics of reaction conditions without
extensive manual feature engineering. To explicitly address concerns of
information asymmetry, we further evaluated Logistic Regression (LR) and
Random Forest (RF) classifiers trained on 1,536-dimensional semantic text
embeddings of the full protocols. While their performance improved significantly,
\OurDatabase still maintained a clear and robust margin, confirming the
necessity of LLM fine-tuning. Nevertheless, the untuned models did preserve
partial ranking structure within some papers, which suggests that chemistry
language models already encode fragments of synthetic plausibility, yet without
the calibration needed to separate a merely feasible solvothermal recipe from
one with credible scalability. Thus, these results provided evidence that
fine-tuning on a literature-derived \OurDatabase dataset containing experimental
knowledge enables the LLM to organize synthesis context into a much sharper
ranking of industrially relevant candidates.

Importantly, these results should be interpreted against the reality that
scalability is not a single property of a framework but a property of a
protocol, and hence the model is assessing procedural plausibility rather than
intrinsic crystallographic ``goodness.'' As supported by our results, often,
scalability is written into solvent choice, reaction severity, and operational
simplicity. Water-based or benign solvent systems, moderate reaction temperatures,
shorter durations, and the absence of complex modulators can correlate with
practical scale-up routes.\cite{ref10,ref43,ref44} Likewise, from an engineering
point of view solvent burden, long solvothermal times, and workup intensity can
be barriers to sustainable manufacturing, which reinforces the idea that these
fields act as weak yet learnable proxies for process viability.\cite{ref11}

In this study, our results demonstrate that such scalability intuition, expressed
neither as a single heuristic nor as a hand-built descriptor set, can become a
learned judgment grounded in literature-scale experimental data. To this regard,
the present study also sheds light on a broader role for large language models
in materials chemistry, wherein they not only extract facts (GPT-5.4 in this
study), but also help estimate whether a new material can move beyond discovery
into practical accessibility (GPT-4.1 and GPT-5.1 in this study). Nevertheless,
it should also be acknowledged that a limitation of the present workflow is that
the input precursor is limited to one in our dataset, therefore it is unable to
predict the scalability of multivariate MOFs that contain more than one metal or
linkers.\cite{ref45}

Furthermore, our definition of a ``scalable'' positive relies on a gram-scale
reporting threshold. The gram-scale threshold represents the minimum quantity at
which a synthesis can meaningfully be discussed as a candidate for process
scale-up, which is sufficient to conduct basic physicochemical characterization,
BET measurement, and preliminary toxicology screening. While our dataset contains
records with 100\,g (Supporting Information, Figure~S2), confirming that
industrial-scale data exist, they remain highly sparse; using 100\,g as the
positive threshold would reduce the training set to $\sim$20 examples, precluding
any data-driven approach. We therefore adopt gram-scale as the lowest
statistically viable threshold, acknowledging that true industrial scalability
requires further validation beyond this literature-based proxy.

Consequently, the set of confirmed successful scale-up protocols remains only in
the hundreds, and thus the positive class, while informative, is still modest
relative to the breadth of MOF chemistry. It is envisioned that, as more
gram-scale, kilogram-scale, and pilot-oriented procedures appear in the
literature, the \OurDatabase dataset should become more chemically diverse and
more sharply calibrated across families of metals, linkers, and solvent systems.

In summary, we show that scale-up potential prediction for MOFs can be
formulated as a literature-grounded positive-unlabeled learning problem and
addressed effectively with LLMs. By constructing the \OurDatabase dataset and
training on small-scale synthesis context, we obtained a model that identifies
scalable MOF candidates with strong gold-benchmark accuracy (93.5\%) and, more
importantly, high deployment ranking power. Conceptually, the study suggests
that synthesis procedures contain latent information about potential future
manufacturability well before industrial development begins. It is therefore
conceivable that, when a new MOF is discovered, one may already estimate
whether its path toward practical scale-up is narrow or promising. Moreover,
for existing reported MOF protocols, this workflow can potentially identify
overlooked ones that contain high scale-up potential and thus commercial
interest. In this regard, this proof-of-concept demonstrates how LLM-driven
ranking tools may become a valuable early filter, helping the field prioritize
which discoveries to evaluate for industrial scale-up.

\section*{Author Contributions}
Z.Z.\ and P.W.\ conceptualized the study and designed the methodology.
Z.Z., P.W., H.S., X.L., and B.F.\ conducted the formal analysis, model
training, and data interpretation. Z.Z., H.S., and X.L.\ were responsible for
the model, workflow, and baseline development. R.C.\ and S.C.P.\ carried out
the experimental validation on zirconium MOFs. Z.Z.\ and H.S.\ prepared the
visualizations. Z.Z., P.W., H.S., and X.L.\ drafted the original manuscript.
All authors engaged in critical review, editing, and provided feedback on the
final manuscript.

\section*{Conflicts of Interest}
There are no conflicts to declare.

\section*{Data Availability}
Data for this article, including text-mined synthesis protocols, structured
reaction parameters, and model-ready JSON datasets, are available at the GitHub
repository at \url{https://github.com/zzhenglab/MOF-Scaleup/}.

\section*{Acknowledgements}
Z.Z.\ acknowledges support from the EQT Foundation Breakthrough Science grant
(No.\ 059994) and the Royal Society of Chemistry through a Researcher
Collaborations grant (GR0042769). Z.Z.\ also appreciates insightful discussions
with Prof.\ Berend Smit from EPFL, Prof.\ Phillip Milner from Cornell
University, and Prof.\ Chenfeng Ke from WashU. S.C.P.\ acknowledges valuable
support and discussions with Prof.\ Banglin Chen.



\clearpage
\newpage

\setcounter{section}{0}
\setcounter{figure}{0}
\setcounter{table}{0}
\setcounter{equation}{0}
\renewcommand{\thesection}{S\arabic{section}}
\renewcommand{\thefigure}{S\arabic{figure}}
\renewcommand{\thetable}{S\arabic{table}}
\renewcommand{\theequation}{S\arabic{equation}}

\begin{center}
  \rule{\textwidth}{0.5pt}\\[0.8em]
  {\LARGE\textbf{Supporting Information}}\\[0.4em]
  {\large Predicting Scale-Up Plausibility of Metal-Organic Framework
  Syntheses}\\[0.4em]
  \rule{\textwidth}{0.5pt}
\end{center}
\vspace{1em}
\tableofcontents
\newpage

\newpage

\section{\OurDatabase Dataset Construction}
\label{sec:si_dataset}

\subsection{Literature Retrieval}
\label{sec:si_retrieval}

The \OurDatabase dataset was assembled from two literature pools retrieved from \textit{Web of Science}.
The \emph{possible-positive pool} (\textbf{P}) was collected using the search keywords ``scale-up'', ``gram-scale'', ``kilogram'', and ``pilot'' within title or abstract fields, restricted to peer-reviewed
publications under the topic ``metal-organic framework'' from 1995 to 2026.
This intentionally broad query yielded 117 candidate paper groups.
The \emph{unlabeled pool} (\textbf{U}) used the keywords ``metal-organic framework'' and ``synthesis'' over the same date range,
then restricted to solvothermal syntheses containing one principal metal source and one principal linker, yielding 946 candidate paper groups.

\subsection{Label Definitions}
\label{sec:si_labels}

Each extracted synthesis protocol was assigned one of four labels (Table~\ref{tab:si_labels}).

\begin{table}[H]
\centering
\caption{Label definitions in the \OurDatabase dataset.}
\label{tab:si_labels}
\small
\begin{tabular}{@{}llp{7.2cm}c@{}}
\toprule
\textbf{Label} & \textbf{Full name} & \textbf{Definition} & \textbf{Training} \\
\midrule
$\mathrm{P_s}$ & Strong positive     & Protocol with explicit scale-up evidence (e.g., gram-scale product mass, multi-gram batch, or pilot-scale report). & Yes $\rightarrow$ ``P'' \\[3pt]
$\mathrm{P_a}$ & Auxiliary positive  & Protocol from the unlabeled corpus that matches a $\mathrm{P_s}$ entry by MOF name and metal/linker identity, representing a small-scale preparation of a MOF later shown to be scalable. & Yes $\rightarrow$ ``P'' \\[3pt]
U               & Unlabeled           & Protocol with no known scale-up evidence; may contain hidden positives. & Yes $\rightarrow$ ``U'' \\[3pt]
N               & Negative            & Expert-curated protocol unlikely to be scalable, identified based on mg-scale-only yields and structural or chemical judgment (e.g., toxic solvents at extreme conditions, single-crystal-only procedures). & \textbf{No} (held out) \\
\bottomrule
\end{tabular}
\end{table}

During training, $\mathrm{P_s}$ and $\mathrm{P_a}$ are both mapped to the positive label ``P'', while U remains ``U''.
Negative examples (N) are completely withheld from training and reserved exclusively for the gold-standard evaluation benchmark.

\newpage
\subsection{Dataset Statistics}
\label{sec:si_stats}

After deduplication and quality filtering, the \OurDatabase dataset comprises
3,568 synthesis protocols.  The label distribution is summarised in
Table~\ref{tab:si_dataset_stats}.

\begin{table}[H]
\centering
\caption{\OurDatabase dataset label distribution.}
\label{tab:si_dataset_stats}
\begin{tabular}{@{}lrr@{}}
\toprule
\textbf{Label} & \textbf{Count} & \textbf{Percentage} \\
\midrule
$\mathrm{P_s}$ (strong positive)    &  379  &  10.6\% \\
$\mathrm{P_a}$ (auxiliary positive) &  344  &   9.6\%  \\
U (unlabeled) & 2,684 & 75.2\% \\
N (negative)  & 161   & 4.5\%  \\
\midrule
\textbf{Total} & \textbf{3,568} & 100\% \\
\bottomrule
\end{tabular}
\end{table}

\begin{figure}[H]
  \centering
  \includegraphics[width=\textwidth, height=0.85\textheight, keepaspectratio]{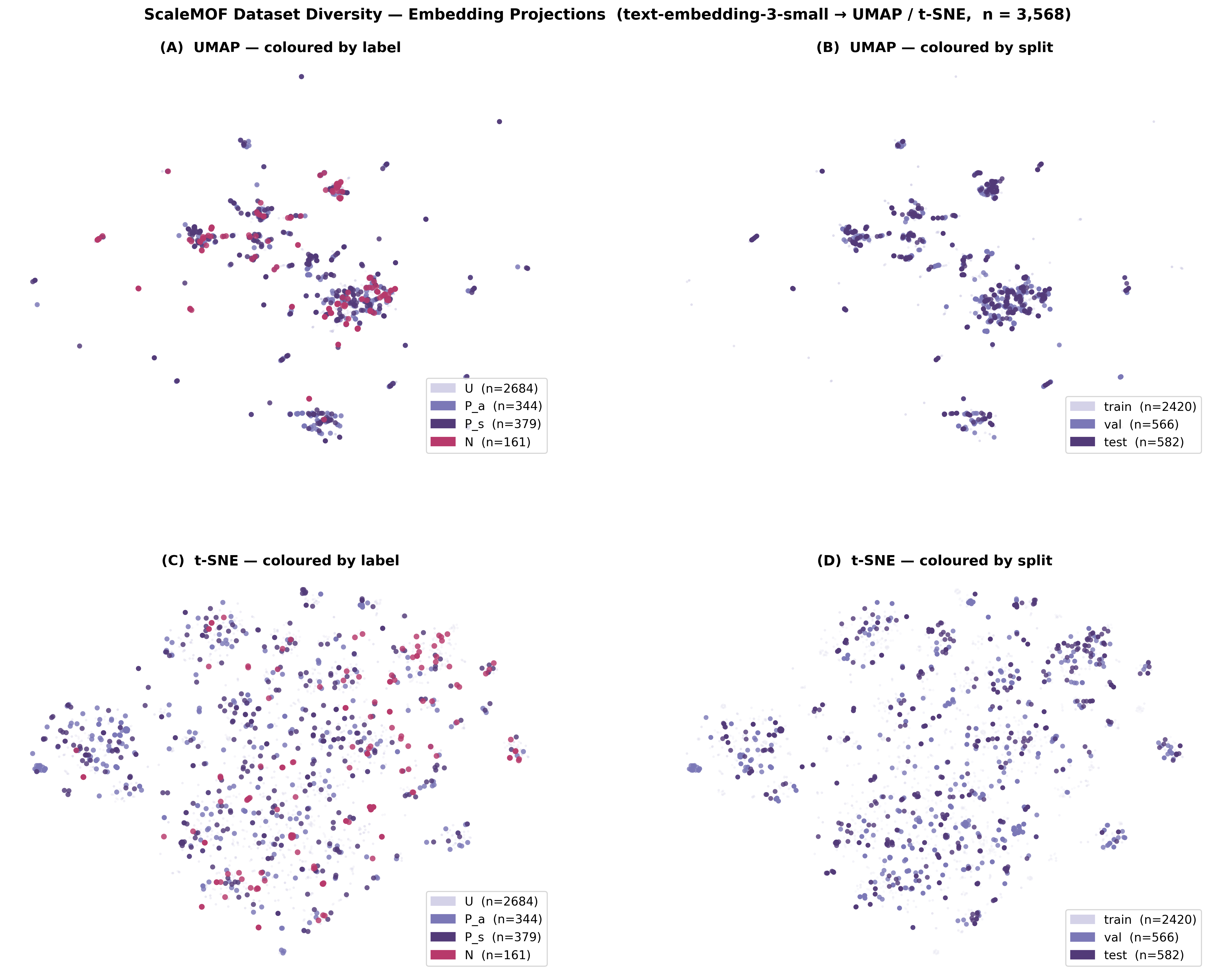}
  \caption{%
    Dataset diversity of the \OurDatabase dataset ($n = 3{,}568$ protocols)
    — embedding projections.
    \textbf{(A)} UMAP and \textbf{(C)} t-SNE projections of 1,536-dimensional
    text embeddings (\texttt{text-embedding-3-small}), coloured by label class
    ($\mathrm{P_s}$: dark purple; $\mathrm{P_a}$: medium purple; U: light purple; N: rose pink).
    \textbf{(B, D)} The same projections coloured by dataset split (train/val/test),
    confirming that no single label cluster is concentrated in one split.
    The broad overlap of all label classes confirms that
    the ScaleMOF dataset samples a chemically diverse region of synthesis-condition
    space rather than a narrow subset of MOF families.
  }
  \label{fig:dataset_diversity}
\end{figure}

\begin{figure}[H]
  \centering
    \includegraphics[width=\textwidth, height=0.80\textheight, keepaspectratio]{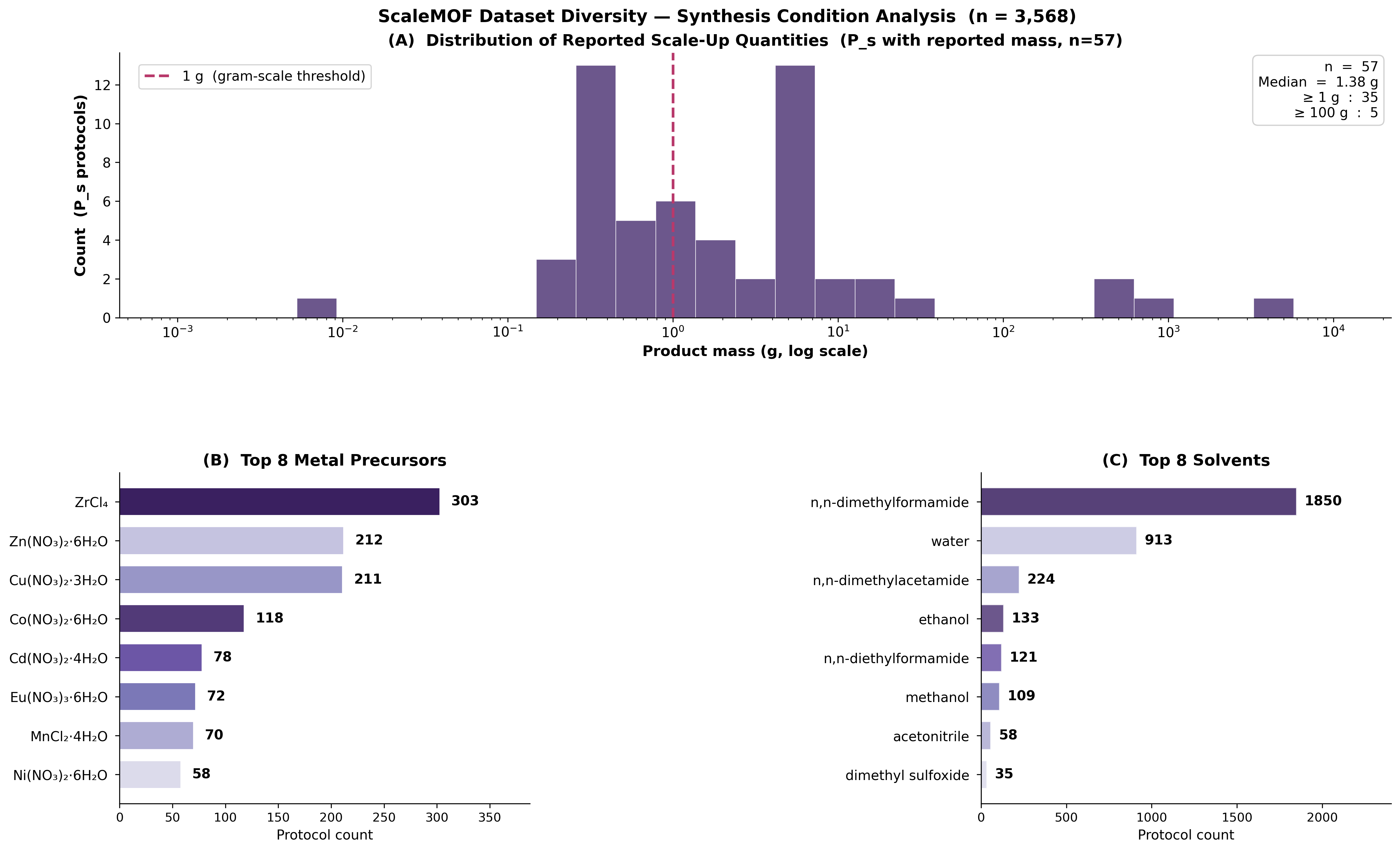}
  \caption{%
    Synthesis condition diversity of the \OurDatabase dataset ($n = 3{,}568$ protocols).
    \textbf{(A)} Log-scale histogram of reported product mass for all $\mathrm{P_s}$ protocols
    with an explicitly recorded product mass ($n=57$);
    the dashed line marks the 1\,g gram-scale threshold used to define the positive class.
    Most records cluster in the 0.1--10\,g range, with a small number reporting industrial-scale
    quantities (\textgreater{}100\,g).
    \textbf{(B, C)} Top 8 metal precursors and top 8 primary solvents by frequency
    across the full dataset, showing the broad coverage of transition metals
    and a variety of solvent systems (aqueous, amide-based, alcoholic).
  }
  \label{fig:dataset_diversity_ef}
\end{figure}

\medskip
\noindent %
\textbf{Dataset Diversity.}
Figures~\ref{fig:dataset_diversity} and~\ref{fig:dataset_diversity_ef} illustrate the chemical
diversity of the 3,568 synthesis protocols.
Figure~\ref{fig:dataset_diversity} presents UMAP and t-SNE projections of 1,536-dimensional text
embeddings, coloured by label class (A, C) and by dataset split (B, D), confirming broad coverage
with no obvious label-correlated cluster separation in embedding space.
Figure~\ref{fig:dataset_diversity_ef} shows the product-mass distribution for $\mathrm{P_s}$ protocols (E)
and the top-8 metal precursors and primary solvents by frequency (F), demonstrating that the
dataset spans diverse metals and solvent systems rather than a narrow chemical subset.

\subsection{Paper-Level Data Splitting}
\label{sec:si_split}

To prevent data leakage, all splits were performed at the \emph{paper-group} level:
every protocol extracted from the same publication (including main text and supporting information) belongs to a single split.  We used stratified splitting based on the dominant label within each paper group, with a 70\,:\,15\,:\,15 ratio for train, validation, and test sets, yielding  644, 174, and 179 paper groups across the three sets, respectively.
The splitting was performed using scikit-learn's \texttt{train\_test\_split} with stratification on the dominant label per paper group.

\medskip
\noindent %
\textbf{Independent allocation of N examples (revision update).}
Negative examples (N, $n=161$) are excluded from the paper-group split and allocated independently:
the full negative set is partitioned into two equal halves (80 for validation, 81 for test) by
a separate stratified shuffle prior to paper-group splitting of the P/U records.
This ensures that all 161 N examples contribute to gold-benchmark evaluation rather than being
partially discarded through paper-group assignment.
Under the previous scheme, 114 of 161 N examples fell into the training paper-group and were
wasted; this correction fully utilises the negative set.

The training split for positive-unlabeled (PU) learning contained
290 $\mathrm{P_s}$, 234 $\mathrm{P_a}$, and 1,896 U protocols (2,420 total).
The evaluation splits are summarised in Table~\ref{tab:splits}.

\begin{table}[H]
\centering
\caption{Evaluation split composition.}
\label{tab:splits}
\begin{tabular}{@{}llccc@{}}
\toprule
\textbf{Benchmark} & \textbf{Split} & \textbf{Positive class} & \textbf{Negative class} & \textbf{$n$} \\
\midrule
Gold ($\mathrm{P_s}$ vs.\ N)     & Validation & 38 $\mathrm{P_s}$ & 80 N & 118  \\
Gold ($\mathrm{P_s}$ vs.\ N)     & Test       & 51 $\mathrm{P_s}$ & 81 N & 132  \\
\addlinespace
Deploy ($\mathrm{P_a}$ vs.\ U) & Validation & 47 $\mathrm{P_a}$ & 401 U & 448 \\
Deploy ($\mathrm{P_a}$ vs.\ U) & Test       & 63 $\mathrm{P_a}$  & 387 U & 450 \\
\bottomrule
\end{tabular}
\end{table}

\newpage
\section{Workflow Implementation Details}
\label{sec:si_workflow}

The workflow consists of seven sequential Steps.
In this section we provide details of each step with reference to the source code modules.
All data and codes are available at \underline{\url{https://github.com/zzhenglab/MOF-Scaleup/}}

\subsection{PDF Text Extraction}
\label{sec:si_pdf}

Full-text PDFs (main manuscripts and supporting information) are parsed into plain text using \texttt{pypdf}.
As most documents exceed the usable context length of the extraction LLM,
we apply a \emph{keyword-guided truncation} scheme to retain passages in synthesis conditions and scale-up details.

The extractor uses a curated set of domain-specific keywords
(e.g., ``solvothermal'', ``autoclave'', ``gram-scale'', ``metal'', ``linker'', ``modulator'', ``yield'', ``scale-up'', ``kilogram'')
as anchors to identify relevant regions.
The extraction procedure consists of the following steps:
\begin{enumerate}[nosep]
\item Scan all pages to locate keyword occurrences.
\item Define context windows around each occurrence.
\item Merge overlapping windows.
\item Extract the merged text regions, subject to a configurable maximum character limit.
\end{enumerate}
This ensures that synthesis procedure sections receive priority over unrelated content (e.g., characterisation data, discussion),
while respecting the token budget of the extraction LLM.
Main text and supporting information are processed independently and then combined before being passed to the extraction agent, enabling cross-referencing between dispersed procedural details.

\paragraph{Keyword selection planner (\texttt{PdfConfigPlanner}).}
The keyword inventory and truncation parameters are stored in \texttt{pdf\_config.json}.
To initialize this configuration, human experts provide a small set of high-priority seed terms.
These seeds are then expanded by a dedicated \texttt{PdfConfigPlanner}, by calling GPT-5.4 to propose additional synthesis-relevant keywords for PDF pre-selection.

The planner takes any existing keywords as input and is instructed to expand them to \texttt{MIN\_FOCUS\_KEYWORDS} distinct entries, spanning experimental section markers, reactor or vessel types, scale-up and quantity indicators, solvent and reaction medium descriptors, work-up terminology, and yield/characterisation language.

Numeric truncation parameters (e.g., window sizes, character limits) are treated as engineering constants.
The planner is invoked at runtime if \texttt{pdf\_config.json} is missing or the keyword count falls below the required threshold, leading consistent coverage without manual intervention.

\medskip
\noindent %
{\sloppy
\textbf{Reproducibility of planner-generated resources.}
All planner-generated resources---\texttt{pdf\_config.json} (focus keywords),
\texttt{solvent\_aliases.json} (solvent canonicalisation table), and versioned prompt files
(\texttt{prompts/extraction\_v1.json}, \texttt{prompts/classification\_v1.json})---are
released in the GitHub repository (\texttt{src/mof\_scaleup/data/} and \texttt{src/mof\_scaleup/prompts/}).
Downstream users can reproduce the full pipeline without re-invoking the planners.
}

\subsection{Structured Extraction}
\label{sec:si_extraction}

An \texttt{ExtractionAgent} class invokes a large language model with schema-constrained output to extract structured synthesis records.
The agent currently supports two API providers:
\begin{itemize}[nosep]
\item \textbf{OpenAI}: Uses the Responses API with structured output mode, where the output schema is enforced by the API itself.
\item \textbf{Anthropic}: Uses the Messages API with JSON output parsing at the application level.
\end{itemize}
Default to OpenAI (GPT-5.4).

\paragraph{Retry-with-feedback mechanism.}
When the LLM output fails Pydantic schema validation, the validation error is appended to the conversation as a user message
(e.g., ``Your output had error: \{error\}. Fix and retry.''), and the model is prompted to self-correct.
This loop runs for up to 2 retries with increasing backoff
($1.0 \times (\text{attempt}+1)$~seconds).
In our practice, the retry mechanism is triggered for approximately $8-10\%$ of papers, and after retries, the overall extraction success rate exceeds $99\%$.

\paragraph{User prompt template.}
Each extraction call receives a user prompt that includes:
(i) the file identifier (DOI or filename),
(ii) the full article text (truncated), and
(iii) the supporting information text (if available),
delimited by explicit start/end markers.
Full prompts see ~\ref{sec:si_prompts}.

\paragraph{Session memory via a content-addressable extraction cache.}
{\sloppy
We introduce a content-addressable session memory, \texttt{ExtractionCache} (\texttt{mof\_scaleup.cache}), to reduce redundant structured-extraction calls, one of the dominant cost in the agent loop.
Unlike the coarse \texttt{-{}-skip-existing} shortcut, the cache keys each episode by its full reasoning context,
\[
\langle \text{provider},\ \text{model},\ \text{prompt\_version},\ \text{file\_id},\ \text{sha256(main\_pdf} \Vert \text{si\_pdf)} \rangle,
\]
with 8-byte length-prefixing before hashing to ensure collision-safe separation between manuscript and SI.
This enforces strict \emph{episodic isolation}:
reuse occurs only when both input bytes and inference configuration match.
On recall, cached outputs are revalidated via \texttt{ArticleExtraction.model\_validate()}.
Any stale or schema-incompatible entry is evicted before falling back to a fresh LLM call.
Successful extractions are written back with full configuration metadata and a UTC timestamp, yielding a replayable audit trail.
The mechanism is opt-in via \texttt{-{}-cache-dir~<path>} and otherwise leaves the pipeline fully stateless.
When enabled, recall costs only a hash over raw PDF bytes, negligible relative to a single model invocation.
The plain-JSON storage design keeps this strategy of low external dependency.
It enables straightforward extensions such as TTL policies, drift-aware invalidation, or long-horizon decision logs.
}

\subsection{Protocol Normalisation}
\label{sec:si_normalisation}

Extracted records undergo three post-processing steps:

\paragraph{(a) Solvent canonicalisation.}
A lookup table of most commonly known solvent name/abbreviation mappings is used to standardise solvent identifiers.
For example, ``DMF'', ``dimethylformamide'', and ``N,N-DMF'' are all mapped to the canonical form ``$N,N$-dimethylformamide'' with abbreviation ``DMF''.
The lookup table uses normalised keys (lowercased, whitespace/punctuation removed, Greek letters transliterated) for robust matching.

\textit{Solvent table Planner} (\texttt{SolventAliasPlanner}).
The alias table is stored in \texttt{solvent\_aliases.json} and maintained by a \texttt{SolventAliasPlanner} class that calls GPT-5.4 to enumerate alias-to-canonical mappings.
Each entry maps a solvent alias (full name, abbreviation, or spelling variant) to a two-element array \lstinline{[canonical_full_name, canonical_abbreviation]}.
The planner augments the existing table to \lstinline{MIN_ALIAS_ENTRIES} entries, covering common MOF solvents
(water, DMF, DEF, DMAc, EtOH, MeOH, DMSO, MeCN, THF, 1,4-dioxane, acetone, NMP, \ldots),
halogenated and aromatic solvents, acid/amine modulators used as co-solvents, fluorinated solvents, and alternate IUPAC names or common misspellings.
Existing entries always take precedence, and the LLM fills gaps only.
The planner is invoked automatically at runtime if the alias file has fewer than \texttt{MIN\_ALIAS\_ENTRIES}.

The solvent system is compacted so that empty slots are removed and single-solvent systems receive a default ratio of~1.

\paragraph{(b) Stoichiometry computation.}
If the metal-to-linker molar ratio is not explicitly reported by the authors,
it is computed from the extracted molar amounts (mmol, mol, or $\mu$mol) of the primary metal source and linker.
The ratio text is formatted in $M\!:\!L$ notation (e.g., ``1:2'').
The stoichiometry source is tracked as \texttt{explicit}, \texttt{computed}, or \texttt{not\_reported}.

\paragraph{(c) Yield parsing.}
If a yield percentage was reported as text (e.g., ``85\%'') but the numeric field was left empty by the extraction,
a regex-based parser extracts the numeric value.

\paragraph{(d) Scale evidence reconciliation.}
If any synthesis record within a paper has a scale label of ``gram'',
``multi-gram'', or ``kilogram'', the article-level flag \texttt{gram\_scale\_reported} is set to ``yes''.
Product masses $\geq 1.0$\,g also trigger this flag.

\subsection{Label Construction and Extraction Schema}
\label{sec:si_schema}

Each synthesis protocol is represented as a \texttt{SynthesisRecord} object conforming to a strict Pydantic schema (\texttt{extra="forbid"}).
The full field hierarchy is described in Table~\ref{tab:schema}.

The \texttt{ScaleEvidence} sub-schema classifies product mass into one of six ordinal scale labels:
\texttt{not\_reported}, \texttt{mg}, \texttt{sub-gram}, \texttt{gram}, \texttt{multi-gram}, and \texttt{kilogram}.

The \texttt{ReportedAmounts} sub-schema captures up to two independently reported quantities per reagent
(e.g., ``0.582\,g'' and ``2.00\,mmol''), preserving the original verbatim text, numeric value, and unit separately.

\begin{table}[H]
\centering
\caption{Fields in the \texttt{SynthesisRecord} extraction schema.}
\label{tab:schema}
\small
\begin{tabular}{@{}llp{6.5cm}@{}}
\toprule
\textbf{Field} & \textbf{Type} & \textbf{Description} \\
\midrule
\texttt{mof\_name}         & string (opt.) & MOF name or identifier \\
\texttt{crystal\_code}     & string (opt.) & CCDC code if reported \\
\texttt{procedure\_text}   & string (opt.) & Short verbatim excerpt of synthesis paragraph \\
\texttt{metal}             & Reagent       & Primary metal precursor (full name, abbreviation, up to two reported amounts with units) \\
\texttt{linker}            & Reagent       & Primary organic linker \\
\texttt{modulator}         & Reagent (opt.) & Primary modulator if present \\
\texttt{other\_reagents}   & list[Reagent] & Additional reagents or additives \\
\texttt{stoichiometry}     & Stoichiometry & Metal:linker molar ratio (text, value, source: explicit/computed/not\_reported) \\
\texttt{solvent\_system}   & SolventSystem & Up to 3 solvents with canonical names, abbreviations, amounts, and volume ratios \\
\texttt{conditions}        & Conditions    & Temperature (\textdegree{}C), time (h), vessel type, stirring mode \\
\texttt{post\_processing}  & PostProcessing & Washing solvent/cycles, activation temperature/time \\
\texttt{yield\_percent}    & float (opt.)  & Isolated yield (\%) \\
\texttt{scale\_evidence}   & ScaleEvidence & Product mass (text, value, unit, grams), scale label, scalability text and rationale \\
\texttt{reference}         & string        & DOI of the source article \\
\bottomrule
\end{tabular}
\end{table}

\subsection{Paper-Level Splitting}
\label{sec:si_splitting_detail}

The dataset is split into train/validation/test at the paper-group level (Section~\ref{sec:si_split}).
Five JSONL files are produced:
\begin{itemize}[nosep]
\item \texttt{train\_pu.jsonl}: PU training data ( 2,420 examples); N excluded.
  Each record is a three-message turn (\texttt{system}~$\rightarrow$~\texttt{user}~$\rightarrow$~\texttt{assistant}).
\item \texttt{val\_gold\_pn.jsonl}: Validation gold ($\mathrm{P_s}$ vs.\ N, $n=118$).
\item \texttt{test\_gold\_pn.jsonl}: Test gold ($\mathrm{P_s}$ vs.\ N, $n=132$).
\item \texttt{val\_deploy\_pu.jsonl}: Validation deployment ($\mathrm{P_a}$ vs.\ U, $n=448$).
\item \texttt{test\_deploy\_pu.jsonl}: Test deployment ($\mathrm{P_a}$ vs.\ U, $n=450$).
\end{itemize}

Each evaluation record additionally stores metadata (\texttt{example\_id}, \texttt{file\_id}, \texttt{group\_id}, \texttt{original\_label}, \texttt{benchmark})
for downstream analysis.

\subsection{LLM Fine-Tuning}
\label{sec:si_finetune}

Fine-tuning was performed via the OpenAI fine-tuning API.
The training JSONL file is uploaded via the Files API, and a fine-tuning job is created on the base model (GPT-4.1).
The job is polled at 30-second intervals until completion.
Key configuration is listed in Table~\ref{tab:finetune}.

\begin{table}[H]
\centering
\caption{Fine-tuning configuration.}
\label{tab:finetune}
\begin{tabular}{@{}l c}
\toprule
\textbf{Parameter} & \textbf{Value} \\
\midrule
Base model                & GPT-4.1 \\
Training examples         &  2,420 \\
Format                    & 3-message turns (system / user / assistant) \\
Assistant output          & Single token (``P'' or ``U'') \\
Temperature (inference)   & 0 \\
Max completion tokens     & 2 \\
Top logprobs returned     & 5 \\
Job polling interval      & 30 s \\
\bottomrule
\end{tabular}
\end{table}

The model ID of the resulting fine-tuned checkpoint \\
(e.g., \texttt{ft:gpt-4.1-2025-04-14:org:scale-up-score:XXXXXXXX})
is recorded and used for all downstream evaluation.

\newpage
\subsection{Calibrated Scoring}
\label{sec:si_scoring}

Referring to \cite{ref28}, the post-processing of fine-tuned model outputs is implemented in the \texttt{ScoringPipeline} class, which provides a fit-once / transform-many interface.

For each input protocol $x$, the pipeline performs three sequential operations.
\begin{enumerate}
    \item Extracts a raw score from the token-level log-probabilities of the two candidate output tokens (``P'' and ``U''), yielding a continuous confidence value in $[0,1]$.
    \item This score is corrected under the PU-learning framework to account for incomplete positive labelling in the literature-derived training set.
    \item The corrected score is optionally calibrated by Platt scaling on the validation gold split, and a decision threshold selected on the same split is used to obtain the final binary prediction.
\end{enumerate}

In addition to the scalar score, the log-probability extraction module records diagnostic metadata, including the generated token sequence, top-$k$ alternative tokens at the classification position, and warnings when the emitted label disagrees with the highest-probability token.
These records are used for debugging and sanity checking, but do not alter the final score.

Mathematical formal expressions of the raw score, PU correction, calibration procedure, and threshold selection are provided in Section~\ref{sec:si_pu_scoring}.

\newpage
\section{Prompts}
\label{sec:si_prompts}

\noindent\emph{
The full prompt text is provided for reproducibility.
It is flexibly optional to follow this methodological argument, and casual readers may skip directly to Section~\ref{sec:si_prompt_versioning}.
}


\subsection{Extraction Agent}
\label{sec:si_extraction_prompt}

The extraction agent receives the following system-level instruction:

\begin{tcolorbox}[title={Extraction: System Prompt}, colframe=gray!60!black, colback=gray!10!white, fonttitle=\bfseries,breakable]
\begin{lstlisting}
You extract structured data for pristine solvothermal or hydrothermal MOF
synthesis procedures from scientific papers.
[... full prompt unchanged --- see original SI ...]
\end{lstlisting}
\end{tcolorbox}

And user instruction for formatting:
\begin{tcolorbox}[title={Extraction User Prompt}, colframe=gray!60!black, colback=gray!10!white, fonttitle=\bfseries,breakable]
\begin{lstlisting}
Context:
You are a MOF chemist extracting procedure-level data for scalable solvothermal
or hydrothermal MOF synthesis.

File identifier:
{file_id}

Full article text:
<<<ARTICLE_START
{article_text}
ARTICLE_END>>>

Supporting Information text:
<<<SI_START
{si_text}
SI_END>>>
\end{lstlisting}
\end{tcolorbox}

\subsection{Classification Prompt}
\label{sec:si_classification_prompt}

The fine-tuned classifier (GPT-4.1) uses the following concise system prompt, identical for training, evaluation, and all zero-shot LLM baselines:

\begin{tcolorbox}[title={Scale-Up PU prediction: System Prompt}, colframe=gray!60!black, colback=gray!10!white, fonttitle=\bfseries]
\begin{lstlisting}
You are an expert in metal-organic framework synthesis and process scale-up
assessment.
You will receive a JSON object that contains only reagents and conditions.
Your task is to determine whether the protocol shows clear potential for
scale-up to gram-scale or larger production.
    P = positive or possible
    U = unknown or unlikely
Output exactly one uppercase character with no spaces: P or U
\end{lstlisting}
\end{tcolorbox}

\subsection{LLM Planner Prompts}
\label{sec:si_planner_prompts}

Two LLM-based planners (GPT-5.4) are used to generate and augment the JSON configuration files consumed by the data-processing pipeline.
Both planners receive a system prompt plus a user prompt that embeds any existing entries so the model only adds genuinely new content.

\subsubsection*{Pdf Configuration Planner}

The \texttt{PdfConfigPlanner} maintains the list of focus keywords used for sliding-window extraction from PDFs (Section~\ref{sec:si_pdf}).

\begin{tcolorbox}[title={PdfConfigPlanner -- system prompt}, colframe=gray!60!black, colback=gray!10!white, fonttitle=\bfseries]
\begin{lstlisting}
You are an expert in metal-organic framework (MOF) synthesis and scientific
literature mining. You produce machine-readable data in strict JSON format.
\end{lstlisting}
\end{tcolorbox}

\begin{tcolorbox}[title={PdfConfigPlanner -- user prompt template}, colframe=gray!60!black, colback=gray!10!white, fonttitle=\bfseries, breakable]
\begin{lstlisting}
Generate >= {target} focus keywords (lowercase, <=6 words each) for locating
MOF synthesis information in academic PDFs via sliding-window extraction.
[... see original for full prompt ...]
\end{lstlisting}
\end{tcolorbox}

\subsubsection*{Solvent Alias Mapping Planner}

The \texttt{SolventAliasPlanner} builds the solvent canonicalisation table
(Section~\ref{sec:si_normalisation}).

\begin{tcolorbox}[title={SolventAliasPlanner -- system prompt}, colframe=gray!60!black, colback=gray!10!white, fonttitle=\bfseries]
\begin{lstlisting}
You are an expert in metal-organic framework (MOF) chemistry and synthesis.
You produce machine-readable data in strict JSON format.
\end{lstlisting}
\end{tcolorbox}

\begin{tcolorbox}[title={SolventAliasPlanner -- user prompt template}, colframe=gray!60!black, colback=gray!10!white, fonttitle=\bfseries, breakable]
\begin{lstlisting}
Generate >= {target} solvent alias entries commonly seen in MOF synthesis
literature.
[... see original for full prompt ...]
\end{lstlisting}
\end{tcolorbox}

\subsection{Versioned Prompt Registry}
\label{sec:si_prompt_versioning}

All system and user prompts consumed by the extraction agent (Section~\ref{sec:si_extraction}) and the classification model (Section~\ref{sec:si_classification_prompt}) are stored as versioned JSON resources under \texttt{src/mof\_scaleup/prompts/}.  Each file follows the naming convention \texttt{\{family\}\_v\{N\}.json} (e.g.,
\texttt{extraction\_v1.json}, \texttt{classification\_v1.json}) and contains the fields listed in Table~\ref{tab:prompt_schema}.

\begin{table}[H]
\centering
\caption{Fields in a versioned prompt JSON file.}
\label{tab:prompt_schema}
\small
\begin{tabular}{@{}l l p{9cm}@{}}
\toprule
\textbf{Field} & \textbf{Type} & \textbf{Description} \\
\midrule
\texttt{name}            & string & Prompt family (\texttt{extraction} or \texttt{classification}) \\
\texttt{version}         & string & Version tag (e.g.\ \texttt{v1}, \texttt{v2}) \\
\texttt{description}     & string & One-line human-readable summary \\
\texttt{system\_prompt}  & string & System-role instruction text \\
\texttt{user\_template}  & string (opt.) & User-message template with \texttt{\{file\_id\}}, \texttt{\{article\_text\}}, \texttt{\{si\_text\}} placeholders (extraction only) \\
\bottomrule
\end{tabular}
\end{table}

A lightweight registry module (\texttt{mof\_scaleup.prompts}) provides two functions:
\texttt{list\_versions(name)}, which returns the sorted list of available version tags for a prompt family;
and
\texttt{load\_prompt(name, version=None)}, which loads the specified version (defaulting to the latest when \texttt{version} is omitted).

\noindent
\textbf{Integration.}
The \texttt{ExtractionAgent} class accepts optional \texttt{system\_prompt} and \texttt{user\_template} constructor arguments.
When these are \texttt{None} (the default), the original hardcoded constants from \texttt{data.prompts} are used, identical to the behaviour reported in the main paper.
On the CLI, passing \texttt{-{}-prompt-version~v2} loads the corresponding JSON file and injects its contents into the agent and the JSONL preparation step  \lstinline{curate.split.prepare_dataset}).

\noindent
\textbf{Community contribution workflow.}
To propose a new prompt variant, a contributor:
(i)~creates a new JSON file (e.g.\ \texttt{extraction\_v2.json}) following
the schema in Table~\ref{tab:prompt_schema};
(ii)~runs the pipeline with \texttt{-{}-prompt-version~v2};
(iii)~compares metrics against the \texttt{v1} baseline on the standard evaluation splits.
No changes to the core pipeline code are required.

\newpage
\section{Baselines and Evaluation}
\label{sec:si_metrics}

\subsection{Baseline Comparison Results}
\label{sec:si_comparison}

\begin{figure}[p]  
  \centering
    \includegraphics[width=\textwidth, height=0.72\textheight, keepaspectratio]{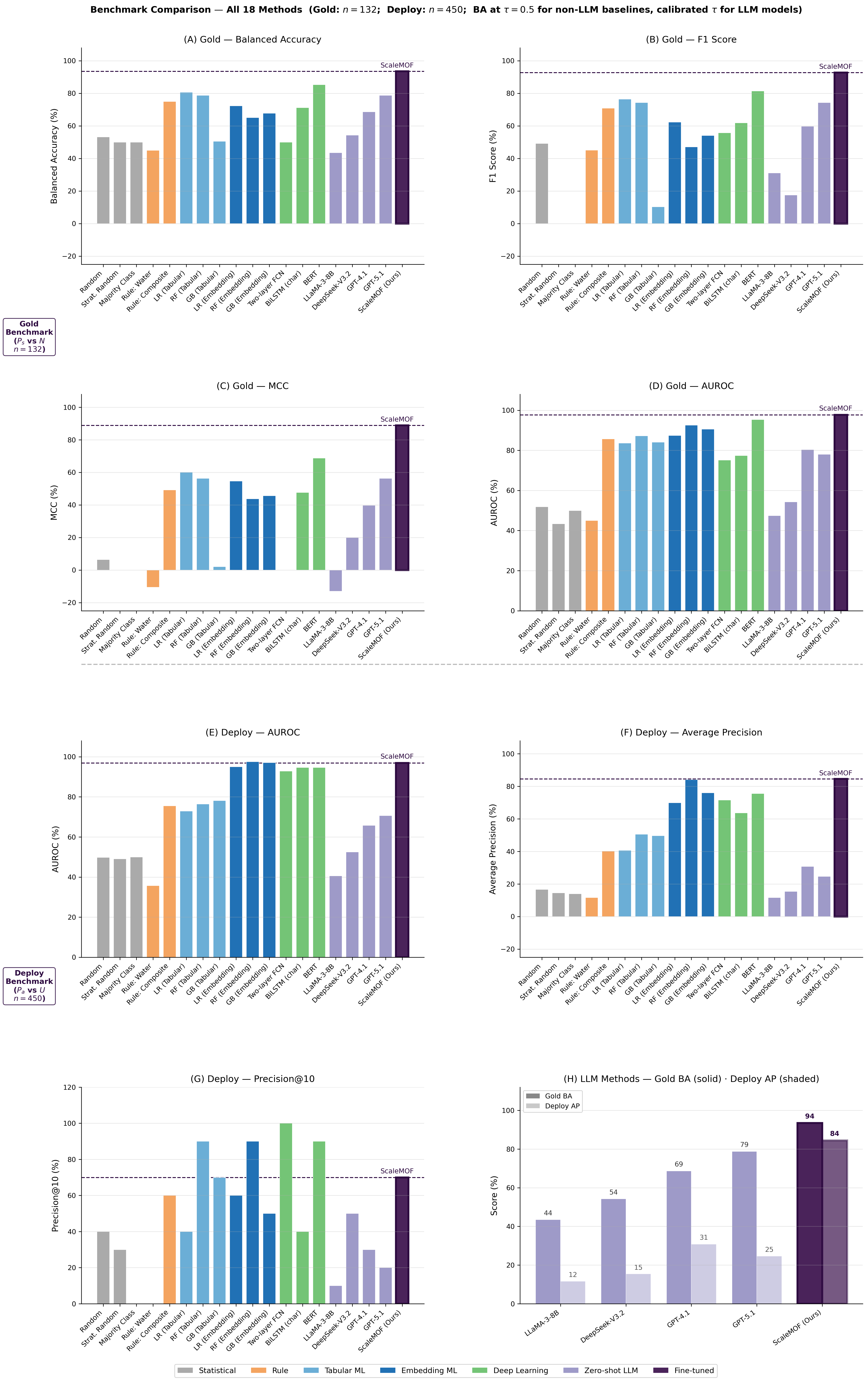}
  \caption{%
    Benchmark comparison across all 18 methods on the held-out test sets
    (4 rows $\times$ 2 columns; panels A--H).
    \textbf{(A, B) Row 1:} Gold benchmark ($\mathrm{P_s}$ vs.\ $N$, $n=132$)
    — Balanced Accuracy and F1 Score.
    \textbf{(C, D) Row 2:} Gold benchmark — MCC and AUROC.
    \textbf{(E, F) Row 3:} Deployment benchmark ($\mathrm{P_a}$ vs.\ $U$, $n=450$)
    — AUROC and Average Precision.
    \textbf{(G, H) Row 4:} Deployment Precision@10 (G) and a focused grouped-bar
    comparison of LLM-based methods showing Gold Balanced Accuracy (solid) and
    Deploy Average Precision (shaded) side by side (H).
    Background shading groups methods by category; the dashed horizontal line
    marks \OurDatabase performance on each metric.
    Balanced accuracy for non-LLM baselines is reported at threshold~$= 0.5$;
    LLM-based models use a threshold calibrated on the validation gold set
    (see Section~\ref{sec:si_scoring}).
    GPT-4.1 gold benchmark evaluated on $n=126$ (6 records excluded due to API failure);
    GPT-4.1 deployment benchmark evaluated on $n=432$ (18 records excluded).
  }
  \label{fig:benchmark_triptych}
\end{figure}

Table~\ref{tab:baselines} summarises all  18 baseline methods on the
held-out test set; Figure~\ref{fig:benchmark_triptych} provides a graphical view.
Per-baseline configuration details are given in Section~\ref{sec:si_baseline_details}.

\begin{table}[H]
\centering
\caption{Comparison of baseline methods with \OurDatabase on the held-out test set. The gold benchmark evaluates classification of explicit scale-up protocols ($\mathrm{P_s}$) against curated negatives (N); the deployment benchmark evaluates ranking of ancillary positives ($\mathrm{P_a}$) within the unlabeled pool (U). Best results per column are \textbf{bolded}.
Balanced accuracy for non-LLM baselines is reported at decision threshold~$= 0.5$; LLM-based models (GPT-4.1, GPT-5.1, ScaleMOF) use a threshold calibrated on the validation gold set (see Section~\ref{sec:si_scoring}).}
\label{tab:baselines}
\begin{subtable}{\textwidth}
\centering
\caption{Gold benchmark ($\mathrm{P_s}$ vs.\ N,  $n=132$)}
\label{tab:baselines_gold}
\begin{tabular}{l c c c c}
\toprule
Method & Bal.\ Acc.\ (\%) & F1 (\%) & MCC (\%) & AUROC (\%) \\
\midrule
  Random                   & 53.2 & 49.2 & 6.4  & 51.9 \\
  Stratified random        & 50.0 & 0.0  & 0.0  & 43.4 \\
  Majority class           & 50.0 & 0.0  & 0.0  & 50.0 \\
\midrule
  Rule: water solvent      & 45.0 & 45.1 & $-$10.6 & 45.0 \\
  Rule: composite heuristic& 75.0 & 70.9 & 49.2 & 85.7 \\
\midrule
  Logistic Regression (Tabular) & 80.7 & 76.4 & 60.1 & 83.6 \\
  Random Forest (Tabular)       & 78.8 & 74.3 & 56.3 & 87.2 \\
  Gradient Boosting (Tabular)   & 50.5$^\dagger$ & 10.3 & 2.1 & 84.1 \\
\midrule
  Logistic Regression (Embedding) & 72.3 & 62.3 & 54.6 & 87.4 \\
  Random Forest (Embedding)       & 65.1 & 47.1 & 43.8 & 92.6 \\
  Gradient Boosting (Embedding)   & 67.8 & 54.1 & 45.6 & 90.6 \\
\midrule
  Two-layer FCN            & 50.0 & 55.7 & 0.0  & 75.1 \\
  BiLSTM (char-level)      & 71.2 & 61.9 & 47.6 & 77.4 \\
  BERT (fine-tuned)        & 85.3 & 81.4 & 68.8 & 95.4 \\
\midrule
  LLaMA-3-8B (zero-shot)   & 43.5 & 31.1 & $-$13.0 & 47.4 \\
  DeepSeek-V3.2 (zero-shot)& 54.3 & 17.5 & 20.0 & 54.3 \\
  GPT-4.1 (zero-shot)      & 68.7$^\ddagger$ & 59.8 & 39.8 & 80.4 \\
  GPT-5.1 (zero-shot)      & 78.8 & 74.3 & 56.3 & 78.0 \\
\midrule
  \textbf{(Ours) ScaleMOF} & \textbf{93.5} & \textbf{92.8} & \textbf{88.9} & \textbf{97.7} \\
\bottomrule
\end{tabular}
\end{subtable}

\vspace{0.5em}
\noindent\small%
$^\dagger$ GB (Tabular) has poorly calibrated output probabilities (mean $\approx 0.25$) at threshold $= 0.5$
because \texttt{GradientBoostingClassifier} does not support \texttt{class\_weight};
AUROC\,=\,84.1\% confirms good ranking ability (see Section~\ref{sec:si_baseline_details}).
$^\ddagger$ GPT-4.1 gold benchmark evaluated on $n=126$ and deployment on $n=432$
(6 and 18 records excluded respectively due to API failures).\par\vspace{0.5em}

\vspace{0.8em}

\begin{subtable}{\textwidth}
\centering
\caption{Deployment benchmark ($\mathrm{P_a}$ vs.\ U,  $n=450$)}
\label{tab:baselines_deploy}
\begin{tabular}{l c c c}
\toprule
Method & AUROC (\%) & AP (\%) & P@10 (\%) \\
\midrule
  Random                   & 49.8 & 16.6 & 40.0 \\
  Stratified random        & 49.1 & 14.5 & 30.0 \\
  Majority class           & 50.0 & 14.0 & -- \\
\midrule
  Rule: water solvent      & 35.7 & 11.7 & 0.0  \\
  Rule: composite heuristic& 75.5 & 40.2 & 60.0 \\
\midrule
  Logistic Regression (Tabular) & 72.9 & 40.6 & 40.0 \\
  Random Forest (Tabular)       & 76.4 & 50.5 & 90.0 \\
  Gradient Boosting (Tabular)   & 78.1 & 49.7 & 70.0 \\
\midrule
  Logistic Regression (Embedding) & 95.0 & 69.9 & 60.0 \\
  Random Forest (Embedding)       & 97.5 & 84.1 & 90.0 \\
  Gradient Boosting (Embedding)   & 97.1 & 76.0 & 50.0 \\
\midrule
  Two-layer FCN            & 92.8 & 71.5 & 100.0 \\
  BiLSTM (char-level)      & 94.6 & 63.6 & 40.0 \\
  BERT (fine-tuned)        & 94.6 & 75.5 & 90.0 \\
\midrule
  LLaMA-3-8B (zero-shot)   & 40.6 & 11.6 & 10.0 \\
  DeepSeek-V3.2 (zero-shot)& 52.5 & 15.4 & 50.0 \\
  GPT-4.1 (zero-shot)      & 65.8$^\ddagger$ & 30.8$^\ddagger$ & 30.0$^\ddagger$ \\
  GPT-5.1 (zero-shot)      & 70.6 & 24.6 & 20.0 \\
\midrule
  \textbf{(Ours) ScaleMOF} & \textbf{96.9} & \textbf{84.5} & \textbf{70.0} \\
\bottomrule
\end{tabular}
\end{subtable}
\end{table}

\begin{table}[H]
\centering
\caption{Paper-level ranking performance for candidate down-selection across  5 held-out test papers in the deployment set.}
\label{tab:paper_level}
\begin{tabular}{l c c c}
\toprule
Method & MAP (\%) & Top-1 Hit (\%) & Top-3 Hit (\%) \\
\midrule
GPT-4.1 (zero-shot) & 75.2 & 80.0 & 100.0 \\
GPT-5.1 (zero-shot) & 70.3 & 40.0 & 100.0 \\
\midrule
\textbf{ScaleMOF (Ours)} & \textbf{87.4} & \textbf{80.0} & \textbf{100.0} \\
\bottomrule
\end{tabular}
\end{table}

\newpage
\subsection{Baseline Method Details}
\label{sec:si_baseline_details}

The  18 baseline methods are organised into  six categories.  All baselines were evaluated on the identical held-out test splits.

\subsubsection*{Group A: Statistical Baselines (No Training Signal)}

Three baselines requiring no training signal, averaged over 100 random seeds to reduce variance:

\begin{enumerate}[label=(\roman*), nosep]
\item \textbf{Random}: Independent Bernoulli(0.5) prediction per example.
\item \textbf{Stratified random}: Bernoulli($p_{\mathrm{train}}$), where
  $p_{\mathrm{train}}$ is the positive-class prevalence in the training set.
\item \textbf{Majority class}: Always predicts the majority training label
  (deterministic).
\end{enumerate}

\subsubsection*{Group B: Rule-Based Baselines (No Fitting)}

Two domain-informed heuristics operating on raw protocol fields:

\begin{enumerate}[label=(\roman*), nosep]
\item \textbf{Water solvent rule}: Predicts ``P'' if the solvent field contains
  ``water''.  Probability assigned as 0.7 (positive) or 0.3 (negative).

\item \textbf{Composite heuristic}: Weighted score across seven features:

\begin{table}[H]
\centering
\begin{tabular}{@{}lc@{}}
\toprule
\textbf{Feature} & \textbf{Weight} \\
\midrule
Water or benign solvent present & $+2$ \\
Temperature $\leq 120$\,\textdegree{}C & $+1$ \\
Reaction time $\leq 24$\,h & $+1$ \\
Stirring present & $+1$ \\
Common scalable metal (Zr, Cu, Zn, Al, Fe, Cr) & $+1$ \\
Common scalable linker (BDC, BTC, imidazole, fumaric acid) & $+1$ \\
No modulator & $+0.5$ \\
\midrule
\textbf{Threshold} & $\geq 3.5$ \\
\bottomrule
\end{tabular}
\end{table}

  Probability is computed as $\min(\max(\text{score}/7.5,\; 0.01),\; 0.99)$.
\end{enumerate}

\subsubsection*{Group C: Supervised Machine Learning (Engineered Features)}

Three scikit-learn classifiers trained on a 14-dimensional feature vector
(Table~\ref{tab:features}).

\begin{table}[H]
\centering
\caption{Engineered features for supervised ML baselines.}
\label{tab:features}
\begin{tabular}{@{}llp{8cm}@{}}
\toprule
\textbf{Feature} & \textbf{Type} & \textbf{Description} \\
\midrule
\texttt{temperature}    & numeric & Reaction temperature (\textdegree{}C), median-imputed \\
\texttt{time\_h}        & numeric & Reaction time (hours), median-imputed \\
\texttt{yield\_pct}     & numeric & Reported yield (\%), median-imputed \\
\texttt{has\_water}     & binary  & Solvent contains water \\
\texttt{has\_dmf}       & binary  & Solvent contains DMF \\
\texttt{has\_ethanol}   & binary  & Solvent contains ethanol \\
\texttt{has\_methanol}  & binary  & Solvent contains methanol \\
\texttt{num\_solvents}  & integer & Count of solvents in the system \\
\texttt{has\_modulator} & binary  & Modulator is present \\
\texttt{has\_stirring}  & binary  & Stirring reported \\
\texttt{has\_yield}     & binary  & Yield explicitly reported \\
\texttt{vessel\_score}  & ordinal & $+2$ (reactor/flask), $+1$ (autoclave), $-1$ (vial), $0$ (other) \\
\texttt{common\_metal}  & binary  & Industrially common metal \\
\texttt{common\_linker} & binary  & Industrially common linker \\
\bottomrule
\end{tabular}
\end{table}

\noindent
Features were standardised with \texttt{StandardScaler}.  The three classifiers
and their hyperparameters:

\begin{enumerate}[label=(\roman*), nosep]
\item \textbf{Logistic Regression}: $C = 0.5$, balanced class weights,
  500 max iterations.
\item \textbf{Random Forest}: 50 trees, max depth 2, balanced class weights.
\item \textbf{Gradient Boosting}: 50 estimators, max depth 2,
  learning rate 0.1.
\end{enumerate}

For the final classifiers, 4 selected features (\texttt{temperature},
\texttt{num\_solvents}, \texttt{has\_modulator}, \texttt{common\_metal})
were used after preliminary analysis.

\noindent\textit{Note on Gradient Boosting calibration.}
\texttt{GradientBoostingClassifier} does not support the \texttt{class\_weight}
parameter; with an imbalanced training set, its \texttt{predict\_proba} outputs
tend to be biased toward the majority class (mean predicted probability $\approx
0.25$ for the positive class).
At threshold~$= 0.5$, GB behaves near-randomly (BA\,$= 50.5\%$), although its
AUROC remains 84.1\%, indicating good rank discrimination.
This is a known limitation of GB on imbalanced data and does not reflect a
model defect; researchers requiring classification performance from GB should
apply a validation-set-calibrated threshold.

\subsubsection*{Group D: Embedding ML (Semantic Text Embeddings)}

\medskip
\noindent %
Three scikit-learn classifiers trained on 1,536-dimensional semantic text
embeddings of linearised protocol strings, generated via the
\texttt{text-embedding-3-small} model (OpenAI API).
The linearisation format is: ``metal: \{M\}; linker: \{L\}; solvent: \{S\}; temperature: \{T\}\,°C; time: \{h\}\,h; vessel: \{V\}; \ldots''.
Embeddings were computed once and cached to \texttt{out/embedding\_results/embeddings\_cache.npz}
to avoid repeated API calls.
Features were standardised with \texttt{StandardScaler(with\_mean=False)} (sparse-safe).

\begin{enumerate}[label=(\roman*), nosep]
\item \textbf{Logistic Regression (Embedding)}: $C = 1.0$, \texttt{solver=``saga''}, 1,000 max iterations.
  (The \texttt{lbfgs} solver is numerically unstable on 1,536-dimensional inputs; \texttt{saga} is used instead.)
\item \textbf{Random Forest (Embedding)}: 300 trees, \texttt{random\_state=42}.
\item \textbf{Gradient Boosting (Embedding)}: 300 estimators, max depth 4,
  learning rate 0.05, subsample 0.8, \texttt{random\_state=42}.
\end{enumerate}

\noindent
These baselines address the concern that information matching (i.e., comparing the plain-text
description of conditions) might suffice for scale-up classification without fine-tuning.
The results (Table~\ref{tab:baselines_gold}) show that Embedding ML achieves AUC up to 92.6\%
on the gold benchmark and AP up to 84.1\% on the deployment benchmark, which is competitive with
other baselines but still meaningfully below ScaleMOF (AUC\,$= 97.7\%$, AP\,$= 84.5\%$),
demonstrating that fine-tuning imparts discrimination beyond generic semantic similarity.

\subsubsection*{Group E:  Deep Learning Baselines}

Two neural models trained on linearised protocol text
(e.g., ``metal: Zinc; linker: BDC; solvent: DMF and water; temperature: 120.0 \textdegree{}C;
time: 24.0 h; vessel: autoclave; stirring: static; yield percent: 85.0\%''):

\begin{enumerate}[label=(\roman*)]
\item \textbf{BiLSTM (character-level)}: Character embedding
  (dim~32)~$\rightarrow$~bidirectional LSTM (hidden~64, 1~layer)~$\rightarrow$~dropout (0.5)~$\rightarrow$~linear classifier.
  cosine annealing schedule, and class-balanced \texttt{BCEWithLogitsLoss} with weighted random sampling.
  Best checkpoint selected by validation AUROC.
  Maximum input length: 512 characters.

\item \textbf{BERT (fine-tuned)}: \texttt{bert-base-uncased} with only the last transformer block unfrozen.
  [CLS] token representation passes through dropout (0.3) and a linear classifier.
  Maximum input length: 256 tokens.
\end{enumerate}

\subsubsection*{Group F:  Zero-Shot LLM Baselines}

Four LLMs evaluated zero-shot (no fine-tuning, no few-shot examples) with the
identical system and user prompts as the fine-tuned \OurDatabase model:

\begin{enumerate}[label=(\roman*), nosep]
\item \textbf{LLaMA-3-8B} (\texttt{Meta-Llama-3-8B-Instruct}): locally
  deployed via HuggingFace Transformers.  Probabilities from next-token
  logits over $\{$P, U$\}$, normalised by their sum.
\item \textbf{DeepSeek-V3.2} (\texttt{deepseek-chat}): cloud API.
\item \textbf{GPT-4.1}: OpenAI API.  Same base architecture as the fine-tuned
  model.
\item \textbf{GPT-5.1}: OpenAI API.  More recent model generation.
\end{enumerate}

All API-based models used temperature\,$= 0$, \texttt{max\_tokens\,=\,2},
and \texttt{top\_logprobs\,=\,5}.
Each model's outputs were processed through an independent \texttt{ScoringPipeline} (PU correction + Platt scaling),
fitted on that model's own validation gold predictions.
The decision threshold for each LLM-based model is selected by grid search (401 points in $[0,1]$) to maximise balanced accuracy on that model's validation gold set.

\subsection{Gold Benchmark (\texorpdfstring{$\mathrm{P_s}$}{Ps} vs.\ N) Metrics}

This benchmark tests whether the model can distinguish protocols with explicit scale-up evidence from expert-curated negatives.

\begin{itemize}[nosep]
\item \textbf{Balanced accuracy}: $\frac{1}{2}(\mathrm{TPR} + \mathrm{TNR})$,
  insensitive to class imbalance.
\item \textbf{F1 score}: Harmonic mean of precision and recall.
\item \textbf{MCC}:
  \[ \frac{\text{TP}\cdot\text{TN} - \text{FP}\cdot\text{FN}}
  {\sqrt{(\text{TP}+\text{FP})(\text{TP}+\text{FN})(\text{TN}+\text{FP})(\text{TN}+\text{FN})}} \]
\item \textbf{AUROC}: Area under the ROC curve.
\item \textbf{PR-AUC}: Area under the precision--recall curve.
\item \textbf{Brier score}: Mean squared error of calibrated probabilities.
\end{itemize}

\subsection{Deployment Benchmark (\texorpdfstring{$\mathrm{P_a}$}{Pa} vs.\ U) Metrics}

This benchmark simulates realistic screening from the general literature.

\paragraph{Pooled metrics:}
\begin{itemize}[nosep]
\item \textbf{AUROC}: Ranking quality over all test protocols.
\item \textbf{Average Precision (AP)}: Summarises ranking with emphasis on the top of the list.
\item \textbf{Precision@$k$} ($k = 10, 25, 50$): Fraction of true positives among the top-$k$ ranked protocols.
\item \textbf{Recall@$k$}: Fraction of all positives captured in the top $k$.
\item \textbf{First positive rank}: Rank of the first true positive.
\end{itemize}

\paragraph{Paper-level metrics:}
\begin{itemize}[nosep]
\item \textbf{Mean Average Precision (MAP)}: Average of per-paper AP values.
\item \textbf{Mean Reciprocal Rank (MRR)}: Average of $1/\text{rank}$ of the first positive per paper.
\item \textbf{Top-$k$ hit rate} ($k = 1, 3, 5$): Fraction of papers where at least one positive appears in the top $k$ protocols.
\end{itemize}

\subsection*{Dual Evaluation Rationale}

The PU framework motivates the two distinct evaluation benchmarks described in Section~\ref{sec:si_metrics}:

\begin{itemize}[nosep]
\item \textbf{Gold benchmark ($\mathrm{P_s}$ vs.\ N):}
  Tests the model's ability to separate protocols with confirmed scale-up evidence from expert-curated negatives.
  Because both labels are ``clean'' (no label noise), this benchmark directly measures classification accuracy.
  The Platt scaling and threshold are fitted on this benchmark's validation split.

\item \textbf{Deployment benchmark ($\mathrm{P_a}$ vs.\ U):}
  Simulates the realistic use case of screening the general literature for latent scalability potential.
  Auxiliary positives ($\mathrm{P_a}$) serve as ground-truth positives, while the unlabeled pool (U) serves as the negative set.
  Because U contains an unknown fraction of true positives,
  the deployment benchmark is evaluated primarily through ranking metrics (AUROC, AP, Precision@$k$)
  rather than threshold-dependent classification metrics,
  since any true positive in U that the model correctly ranks highly would be penalised as a ``false positive'' under threshold-based metrics.
\end{itemize}

\noindent
This dual-benchmark design allows us to separately assess the model's discriminative power (gold) and its practical utility for literature-scale screening (deployment).

\subsection*{Confusion Matrices}
\noindent
The confusion matrix for the deployment benchmark uses the same decision threshold
$\tau = 0.11$.
Note that the unlabelled pool $U$ may contain an unknown fraction of genuinely scalable
protocols; records classified as ``true negative'' in panel (b) are therefore protocols
ranked below the threshold rather than confirmed negatives.

\begin{figure}[H]
  \centering
    \includegraphics[width=\textwidth, height=0.55\textheight, keepaspectratio]{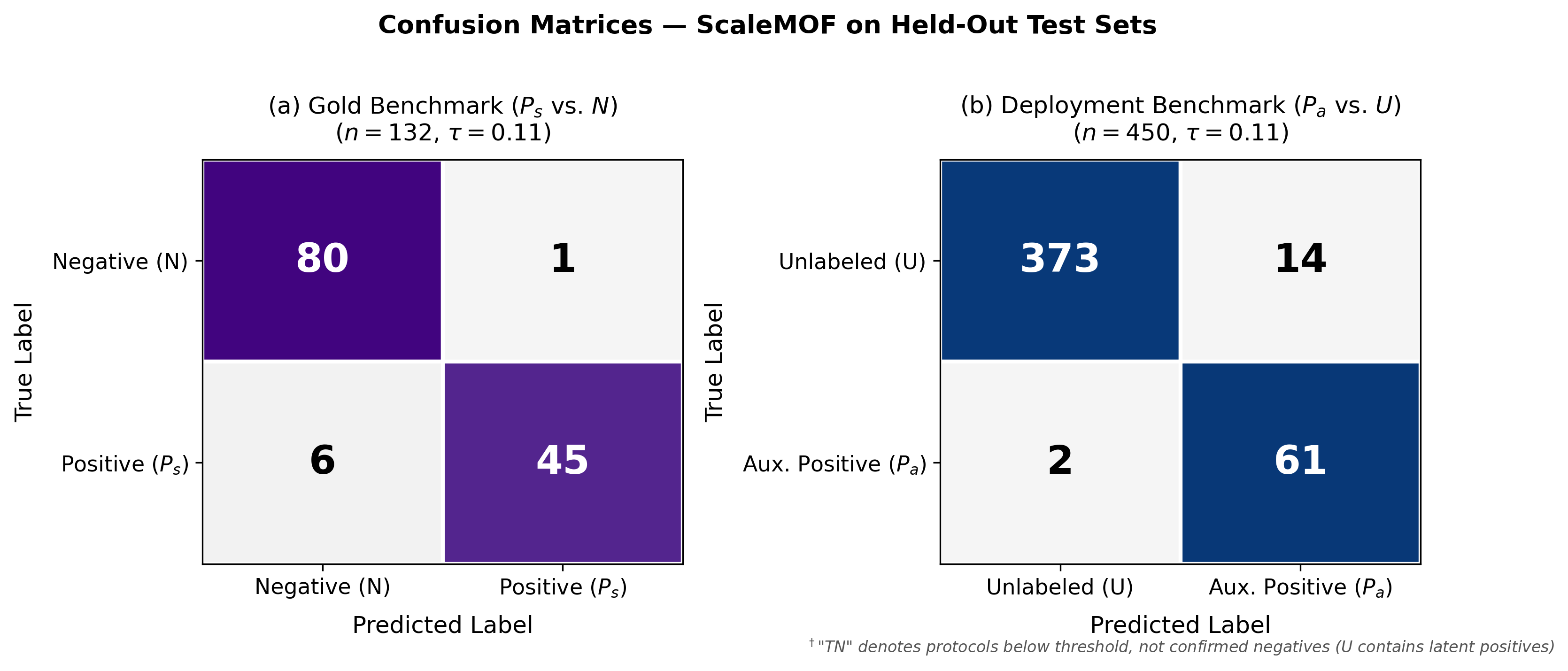}
  \caption{%
    Confusion matrices for the \OurDatabase model on the held-out test sets at
    decision threshold $\tau = 0.11$ (calibrated on the validation gold set).
    \textbf{(a) Gold benchmark} ($\mathrm{P_s}$ vs.\ $N$, $n = 132$):
    TP\,=\,45, TN\,=\,80, FP\,=\,1, FN\,=\,6
    (balanced accuracy 93.5\%, F1 92.8\%).
    \textbf{(b) Deployment benchmark} ($\mathrm{P_a}$ vs.\ $U$, $n = 450$):
    TP\,=\,61, TN\,=\,373, FP\,=\,14, FN\,=\,2.
    Colour encodes row-normalised proportions (recall perspective).
  }
  \label{fig:confusion_matrices}
\end{figure}

\subsection{Extraction Quality Validation}
\label{sec:si_validation}

A random 10\% of selected papers from both the scale-up and general synthesis corpora were manually verified
by comparing LLM-extracted records with ground-truth annotations prepared by a human expert.
The extraction accuracy for key reaction parameters (precursor identity, solvent system, reaction time, and temperature) was 97.6\%,
consistent with reported accuracies for LLM-driven chemical data mining in the literature.

The extraction agent's retry-with-feedback mechanism was triggered in approximately 8\% of papers.
In these cases, Pydantic schema validation errors (typically missing required fields or type mismatches) were appended to the
conversation, allowing the model to self-correct.
After up to 2 retries, the overall extraction success rate exceeded 99\%.

\newpage
\subsection{Bootstrap Confidence Intervals}

Confidence intervals (95\%) were computed using non-parametric bootstrap resampling with 1,000 iterations (seed\,=\,7).
For each iteration, the test set was resampled with replacement, and gold metrics were recomputed.
The 2.5th and 97.5th percentiles define the confidence bounds.

\begin{table}[H]
\centering
\caption{95\% bootstrap confidence intervals for the fine-tuned \OurDatabase model on the gold test set
($n = 132$, 1,000 resamples).}
\label{tab:bootstrap}
\begin{tabular}{@{}lcc@{}}
\toprule
\textbf{Metric} & \textbf{Point estimate} & \textbf{95\% CI} \\
\midrule
Balanced accuracy & 93.5\% & [89.1\%, 97.6\%] \\
F1                & 92.8\% & [87.4\%, 97.4\%] \\
MCC     & 88.9\% & [81.2\%, 95.5\%] \\
AUROC   & 97.7\% & [93.9\%, 100.0\%] \\
\bottomrule
\end{tabular}
\end{table}

\subsection{Evaluation Inference Infrastructure}
\label{sec:si_runner}

Model outputs on all evaluation splits were collected via an asynchronous
batch runner (\texttt{AsyncEvalRunner}) shared across the fine-tuned \OurDatabase
model and all zero-shot API baselines.  The runner manages concurrent
inference with:

\begin{itemize}[nosep]
\item Semaphore-based concurrency control (default: 40 simultaneous requests).
\item Exponential backoff retry logic (up to 6 attempts per request).
\item Progress reporting every 20 records.
\item Per-split CSV output for checkpointing.
\end{itemize}

Raw predictions (token-level log-probabilities and text output) are written
to CSV files, which are then passed into the \texttt{ScoringPipeline}
(Section~\ref{sec:si_scoring}) to produce calibrated final scores.  This
separation of inference and calibration allows the pipeline to be fitted once
on the validation set and applied consistently across all test splits.

\newpage
\section{Positive-Unlabeled (PU) Learning Framework}
\label{sec:si_pu_theory}

\subsection{Motivation}

Standard binary classification requires both positive and negative training labels.
In the MOF scale-up prediction task, however, the absence of scale-up evidence in a publication does not imply that the synthesis is fundamentally non-scalable---the authors may simply not have attempted or reported a larger batch.
This \emph{label asymmetry} means that the unlabeled pool $\mathcal{U}$ is a mixture of true negatives and latent positives (i.e., genuinely scalable protocols that lack published scale-up data).
Training a standard classifier on such data would systematically underestimate the scalable class and produce biased probability estimates.

Positive-Unlabeled (PU) learning\cite{ref24} addresses precisely this setting:
the learner has access to a set of verified positives and a larger set of unlabeled examples, but no confirmed negatives.
The framework provides a principled correction that recovers calibrated posterior probabilities from a model trained on $\{\text{P}, \text{U}\}$ labels alone.

\medskip
\noindent %
\textbf{Pragmatic interpretation of the PU framework (revision note).}
We acknowledge that the composite positive set $\mathrm{P} = \mathrm{P_s} \cup \mathrm{P_a}$ does not
strictly satisfy the ``Selected Completely At Random'' (SCAR) assumption.
$\mathrm{P_a}$ was constructed by MOF-identity matching rather than independent random sampling, which may
introduce mild publication bias.
We therefore treat $\hat{c}$ as a pragmatic calibration factor rather than a theoretically exact PU constant.
Estimation uses only $\mathrm{P_s}$ examples (the most reliable positive evidence), providing a stable
lower bound on the true labelling frequency.
The consistently strong performance on both the gold holdout and the deployment benchmark validates this
pragmatic approach empirically.

\subsection{Problem Formulation}

Let $f : \mathcal{X} \to \{0,1\}$ be the true (unknown) binary label,
where $f(x) = 1$ denotes a genuinely scalable protocol and $\mathcal{X}$ is the space of synthesis protocol descriptors (metal, linker, solvent, conditions, etc.).
Let $s(x) \in \{0,1\}$ denote whether the protocol was \emph{selected} (i.e., reported with explicit scale-up evidence in the literature).
In the PU setting the learner observes:
\begin{itemize}[nosep]
\item A set of \emph{labelled positives} $\mathcal{P} = \{x : s(x) = 1\}$, drawn from
  $p(x \mid f(x) = 1, s(x) = 1)$.
  In our dataset, $\mathcal{P}$ comprises both strong positives ($\mathrm{P_s}$, protocols with direct scale-up evidence)
  and auxiliary positives ($\mathrm{P_a}$, small-batch preparations of MOFs later shown to be scalable).
\item A set of \emph{unlabeled examples} $\mathcal{U} = \{x : s(x) = 0\}$, drawn from the marginal $p(x)$,
  which contains an unknown fraction of latent positives alongside true negatives.
\end{itemize}

\noindent
\textbf{Training label mapping.}
During training, both $\mathrm{P_s}$ and $\mathrm{P_a}$ are mapped to the positive label ``P'',
while U examples retain the label ``U''.
Crucially, the expert-curated negative set (N, $n = 161$) is \emph{entirely excluded} from training
and reserved for the gold evaluation benchmark (Section~\ref{sec:si_metrics}),
ensuring that the model is trained under the strict PU assumption with no access to confirmed negatives.
The resulting training set contains 
2,420 examples (290 $\mathrm{P_s}$ + 234 $\mathrm{P_a}$ + 1,896 U).

\subsection{The SCAR Assumption}

The key Selected Completely At Random (SCAR) assumption\cite{ref24} states that, among truly scalable protocols,
the probability of being labelled (i.e., reported with scale-up evidence) is a constant $c$ independent of the protocol features:
\begin{equation}
  p(s = 1 \mid x, f = 1) = c \in (0, 1]
  \label{eq:scar}
\end{equation}
This means that the labelled positives $\mathcal{P}$ are an unbiased sample of all true positives---which ones get reported with scale-up data is random with respect to the synthesis conditions themselves.

\paragraph{Justification in the MOF context.}
Whether a research group reports scale-up results is driven primarily by factors orthogonal to the synthesis conditions themselves---funding priorities, application targets, journal scope, and industrial collaborations---rather than by the specific metal/linker/solvent/temperature combination.
A protocol using Zr/BDC in DMF at 120\,\textdegree{}C is not inherently more or less likely to \emph{appear}
in a scale-up paper than one using Cu/BTC in water at 80\,\textdegree{}C,
even though both may be equally scalable.
While SCAR is an idealisation (publication bias may introduce mild deviations),
it provides a tractable and well-studied framework that is substantially more appropriate
than the naive assumption that all unlabeled protocols are negative.

\paragraph{PU posterior correction.}
Under SCAR, the true positive posterior can be recovered from the observable quantity $p(s = 1 \mid x)$:
\begin{equation}
  p(f = 1 \mid x) = \frac{p(s = 1 \mid x)}{c}
  \label{eq:pu_corrected}
\end{equation}
Since a standard classifier trained on $\{\text{P}, \text{U}\}$ labels learns to approximate $p(s = 1 \mid x)$ rather than $p(f = 1 \mid x)$,
dividing by $c$ corrects for the fraction of true positives that remain hidden in the unlabeled pool.

\subsection{Estimation of the Label Frequency \texorpdfstring{$\hat{c}$}{c-hat}}

The constant $c$---the probability that a truly scalable protocol appears in the labelled positive set---must be estimated from data.
Following the classical estimator of Elkan and Noto,\cite{ref24} we compute $\hat{c}$ as the average raw model confidence
on a held-out set of known positives:
\begin{equation}
  \hat{c} = \frac{1}{|\mathcal{V}_{P_s}|}
  \sum_{x \in \mathcal{V}_{P_s}} q(x)
  \label{eq:c_hat_detail}
\end{equation}
where $q(x)$ is the raw score from the fine-tuned model (Eq.~\ref{eq:raw_score}) and $\mathcal{V}_{P_s}$
is the validation gold set of strong positives ($n = 38$).
This estimator is consistent under SCAR: on true positives the model should assign high confidence,
and the average confidence converges to $c$.
The computed value is clipped to $[10^{-6},\; 0.999]$ for numerical stability.

Our estimated value $\hat{c} = 0.8410$ admits a concrete interpretation:
the fine-tuned model assigns, on average, a raw P-probability of 84.1\% to protocols with confirmed scale-up evidence.
This indicates that (i) the model has learned a strong signal for scalability,
and (ii) the labelled literature captures most but not all genuinely scalable protocols---approximately 16\% of truly scalable MOF syntheses remain hidden as latent positives in the unlabeled pool.
This motivates the PU correction: without it, the model would systematically underestimate scalability probabilities
by a factor of $\sim$1.2$\times$.

\subsection{Three-Stage scalability Scoring Method}
\label{sec:si_pu_scoring}

The fine-tuned model output is converted into a calibrated scalability score by a three-step PU-aware post-processing procedure.
Implementation details see ~\ref{sec:si_scoring}.

\paragraph{Step 1: Raw score extraction.}
For each input protocol $x$, the model generates a single classification token, either ``P'' or ``U''. The raw score $q(x)$ is computed from the token-level log-probabilities:
\begin{equation}
  q(x) = \frac{\exp(\ell_{\mathrm{P}})}
               {\exp(\ell_{\mathrm{P}}) + \exp(\ell_{\mathrm{U}})}
  \label{eq:raw_score}
\end{equation}
where $\ell_{\mathrm{P}}$ and $\ell_{\mathrm{U}}$ denote the log-probabilities of the ``P'' and ``U'' tokens at the classification position. In implementation, this quantity is evaluated using a numerically stable two-class softmax by subtracting $m=\max(\ell_{\mathrm{P}}, \ell_{\mathrm{U}})$ before exponentiation. The resulting $q(x)\in[0,1]$ is interpreted as the model estimate of $p(s=1\mid x)$, where $s=1$ denotes assignment to the observed positive class.

\paragraph{Step 2: PU correction.}
Under the classical PU-learning perspective,\cite{ref24} the observed positive label is only assigned to a subset of truly positive examples. Let $c=p(s=1\mid y=1)$ denote the positive-labelling frequency. We estimate this quantity from the validation $\mathrm{P_s}$ examples as
\begin{equation}
  \hat{c} = \frac{1}{|\mathcal{V}_{P_s}|}
             \sum_{x \in \mathcal{V}_{P_s}} q(x),
  \label{eq:c_hat}
\end{equation}
and obtain the PU-corrected score by
\begin{equation}
  s_{\mathrm{PU}}(x) = \min\!\left(\frac{q(x)}{\hat{c}},\, 1\right).
  \label{eq:pu_correction}
\end{equation}
In our experiments, $\hat{c}=0.8410$ (clipped to $[10^{-6},\,0.999]$ for numerical stability). This correction compensates for the dilution of the positive class by unlabeled-but-potentially-positive examples and inflates intermediate raw scores accordingly.

\paragraph{Step 3: Probability calibration and threshold selection.}
A logistic regression model (scikit-learn \lstinline{LogisticRegression} with L-BFGS solver) is fitted on the PU-corrected scores of the validation gold set ($\mathrm{P_s}$ vs.\ N), producing the final calibrated score
\begin{equation}
  s_{\mathrm{final}}(x) = \sigma\!\left(w\, s_{\mathrm{PU}}(x) + b\right),
  \label{eq:platt}
\end{equation}
where $\sigma$ is the logistic sigmoid and $w,b$ are learned parameters.\cite{ref40} Platt scaling is applied only when both classes are present and the validation set contains at least 10 examples; otherwise, $s_{\mathrm{PU}}(x)$ is used directly as the final score. Before fitting, input scores are clipped to $[10^{-6},\,1-10^{-6}]$ to avoid numerical instability near the boundaries.

The classification threshold is selected by exhaustive grid search over 401 equally spaced points in $[0,1]$, maximizing balanced accuracy on the validation gold split. The resulting threshold is $\tau=0.11$. Its relatively low value reflects the task preference for recovering potentially scalable protocols at high recall, provided that false-positive rates remain acceptable.

\subsection{Score Column Semantics}

Each scored protocol carries three progressively refined probability estimates,
corresponding to the three pipeline stages:

\begin{table}[H]
\centering
\caption{Score columns produced by the \texttt{ScoringPipeline}.}
\label{tab:score_columns}
\begin{tabular}{@{}llp{7cm}@{}}
\toprule
\textbf{Column} & \textbf{Range} & \textbf{Interpretation} \\
\midrule
\texttt{raw\_q}        & $[0, 1]$ & Raw softmax probability $q(x) = p(s{=}1 \mid x)$ from token logprobs. Estimates the probability of being \emph{labelled} positive. \\[3pt]
\texttt{pu\_score}     & $[0, 1]$ & PU-corrected score $q(x) / \hat{c}$. Estimates the probability of being \emph{truly} scalable ($p(f{=}1 \mid x)$). \\[3pt]
\texttt{final\_score}  & $[0, 1]$ & Platt-scaled calibrated probability (or \texttt{pu\_score} if Platt is disabled). Used for threshold-based classification and ranking. \\
\bottomrule
\end{tabular}
\end{table}

\newpage
\section{Software and Reproducibility Resources}
\label{sec:si_reproducibility}

\subsection{Software Dependencies}

\begin{table}[H]
\centering
\caption{Key software dependencies.}
\label{tab:software}
\begin{tabular}{@{}ll@{}}
\toprule
\textbf{Component} & \textbf{Library / Service} \\
\midrule
LLM extraction        & OpenAI API (GPT-5.4) / Anthropic API (Claude) \\
Schema validation     & Pydantic v2 \\
Fine-tuning           & OpenAI Fine-tuning API (GPT-4.1 base) \\
Classical ML          & scikit-learn \\
Embedding ML& OpenAI Embeddings API (\texttt{text-embedding-3-small}), scikit-learn \\
Deep learning         & PyTorch, HuggingFace Transformers \\
Data processing       & pandas, NumPy \\
PDF parsing           & pypdf \\
Async inference       & asyncio, OpenAI AsyncClient \\
\bottomrule
\end{tabular}
\end{table}

\subsection{Key Constants for Reproducibility}

\begin{table}[H]
\centering
\caption{Key hyperparameters and constants.}
\label{tab:constants}
\begin{tabular}{@{}lll@{}}
\toprule
\textbf{Parameter} & \textbf{Value} & \textbf{Purpose} \\
\midrule
Training fraction       & 70\%   & Paper-level allocation \\
Validation fraction     & 15\%   & Paper-level allocation \\
Test fraction           & 15\%   & Paper-level allocation \\
Max retries (extraction)& 2      & LLM extraction retry count \\
Bootstrap resamples     & 1,000  & Confidence interval estimation \\
Bootstrap seed          & 7      & Bootstrap reproducibility \\
Threshold grid          & 401 points & Balanced accuracy optimisation \\
Eval concurrency        & 40     & Async API inference limit \\
API retries (eval)      & 6      & Exponential backoff attempts \\
PU correction $\hat{c}$ & 0.8410 & Estimated from validation $\mathrm{P_s}$ \\
Platt threshold $\tau$  & 0.11 & Maximises balanced accuracy on val gold \\
\bottomrule
\end{tabular}
\end{table}


\end{document}